\documentclass[aps,prb,preprint,superscriptaddress,showpacs]{revtex4} 
\usepackage{graphicx,amsmath}
\usepackage{bm} 
\input{comment.sty}
\newcommand \bea {\begin{eqnarray}} 
\newcommand \eea {\end{eqnarray}} 
\newcommand \ba {\begin{eqnarray*}} 
\newcommand \ea {\end{eqnarray*}} 
\newcommand \be {\begin{equation}} 
\newcommand \ee {\end{equation}} 
\newcommand{\ka}{{\bm{{\rm k}}}}
\newcommand{\kap}{{\bm{\kappa}}}

\newcommand{\q}{\bm{q}}
\newcommand{\rr}{\bm{r}}
\newcommand{\AAA}{\bm{A}}
\newcommand{\bnabla}{\boldsymbol{\nabla}}
\newcommand{\J}{\bm{J}}
\newcommand{\dagga}{{\phantom{\dagger}}}
\begin{document} 
\title{Role of the impurity-potential range in 
disordered $d$-wave superconductors} 
\author{Luca Dell'Anna}
\affiliation{Institut f\"ur Theoretische Physik, Heinrich-Heine-Universit\"at, 
D-40225 D\"usseldorf, Germany}
\author{Michele Fabrizio}  
\affiliation{International School for Advanced Studies (SISSA), Via Beirut 2-4,  
I-34014 Trieste, Italy}  
\affiliation{The Abdus Salam International Centre for Theoretical Physics  
(ICTP),  
P.O.Box 586, I-34014 Trieste, Italy}  
\author{Claudio Castellani}  
\affiliation{Dipartimento di Fisica, Universit\`a di Roma "La Sapienza", 
and INFM-SMC, Piazzale Aldo Moro 2, I-00185 Roma, Italy}  
\date{\today}  
\begin{abstract}  
We analyze how the range of disorder affects the   
localization properties of quasiparticles in a two-dimensional
$d$-wave superconductor within the standard non-linear $\sigma$-model
approach to disordered systems.  We show that for purely long-range disorder, which 
only induces intra-node scattering processes, the approach is free from the ambiguities
which often beset the disordered Dirac-fermion theories, and
gives rise to a Wess-Zumino-Novikov-Witten action leading to vanishing density of states and  
finite conductivities. We also study the crossover induced by internode scattering 
due to a short range component of the disorder, thus providing  
a coherent non-linear $\sigma$-model description in agreement with all 
the various findings of different approaches.

\end{abstract}  
\maketitle
  
\section{Introduction}
The low-temperature quasiparticle transport in two-dimensional $d$-wave 
superconductors like cuprates is a fascinating issue 
due to the presence of four nodes in the energy spectrum of the Bogoliubov 
quasiparticles, around which the low-lying excitations have a Dirac-like dispersion.
Within the self-consistent Coherent-Phase-Approximation in the limit of weak disorder, spin
and thermal conductivities are found to be related by a Wiedemann-Franz 
law and to acquire universal values which do not depend on the disorder strength.~\cite{Lee} 
Inclusion of quantum interference in the framework of 
the standard non-linear $\sigma$-model  approach to disordered systems\cite{Wegner,EL&K} 
leads to a variety of universality classes\cite{Altland,Fisher,Fisher2,Fukui,yash,Luca,atkinson,all}. 
In the ``generic'' case of short-ranged non magnetic impurities 
full localization of Bogoliubov quasiparticles is predicted.~\cite{Fisher}

Nevertheless, with the only exceptions of YBCO(124)\cite{hussey} and 
Pr$_{2-x}$Ce$_x$CuO$_4$\cite{hill}, experiments in cuprates materials like
YBCO(123)\cite{taillefer,chiao,sutherland}, BSCCO(2212)\cite{chiao} and 
LSCO\cite{takeya} do not show any evidence of strong or even weak localization 
in the superconducting phase down to 0.1 Kelvin.
Various physical effects may be invoked to explain 
the disagreement between theory and experiments.

For instance one may argue that the origin of the discrepancy are spin-flip scattering events,  
even though the systems are nominally free from magnetic impurities. Indeed, in the presence
of spin-flips, the non-linear $\sigma$-model predicts that quantum interference has 
a delocalizing effect\cite{Luca}. Alternatively, or in addition, strong dephasing processes
might set the temperature scale for the onset of localization effects
below the experimentally accessed region. This question has not been settled yet\cite{Bruno}. 
Residual interactions among quasiparticles can also favor 
delocalization\cite{Luca,all}, even though, in the weak disorder limit,
they are expected to be less effective since their 
coupling is proportional to the density 
of states which, already in the Born approximation, is very small.

Another possible explanation invokes the range of the impurity potential.
In the case of purely long range disorder, forward processes dominate and
scattering occurs mainly within each node. 
In the extreme case of intra-node scattering only, it has been shown~\cite{Nersesyan,Zirnbauer} 
that the density of states behaves quite differently from the 
isotropic-scattering case. In addition the eigenstates have been argued\cite{Zirnbauer} 
to be delocalized, unlike for short-range impurity potential. Even though real disorder 
will always have an isotropic component which provides scattering among all four nodes,
and eventually drive the system to localization, one might argue that 
a large value of the intranode with respect to the internode scattering could lower 
the crossover temperature for the appearance of localization precursor effects.
A sizeable amount of long range disorder has been indeed argued to be present in cuprates
on the basis of STM and microwave conductivity experiments\cite{scalapino}. 
This is not surprising since superconducting cuprates are intrinsically disordered by 
the out-of-plane charge dopants which mainly provide a long-range disordered potential. 
Further doping with iso-valent impurities which substitute in-plane Cu-ions 
only adds a short-range component on top of the pre-existing long-range tail of 
the disordered potential.

\medskip 

The results in the presence of purely intra-node impurity scattering 
have been obtained\cite{Nersesyan,Zirnbauer} within 
an approach which is conceptually different from the standard non-linear $\sigma$-model 
approach to disordered systems. The latter starts from the Born approximation, i.e. 
from impurity-damped quasiparticles, and treats perturbatively what is beyond that, i.e. 
quantum interference effects on the diffusive motion. 
On the contrary, the alternative methods used in 
Refs.~[\onlinecite{Nersesyan,Zirnbauer}] do not rely on the Born approximation  
but just map the action of ballistic nodal-quasiparticles in the presence of disorder 
onto the action of one-dimensional (1d) fermions in the presence of an interaction, which  
is generated by the disorder average. Within the mapping, one of the two spatial dimensions  
transforms into the time coordinate of the 1d model while the other into the 1d spatial 
coordinate. The final model is then analyzed by abelian and non-abelian bosonization.  
The outcome of this analysis is that for purely intra-node or inter-opposite-node disorder,
where essentially exact results can be obtained\cite{Nersesyan,Zirnbauer}, 
the density of states is power-law vanishing at the chemical potential with 
an exponent which is disorder-dependent in the former case and universal in the latter.
In both cases the results suggest that a diffusive quasiparticle motion never sets in, namely quasiparticles move 
ballistically down to zero energy.  
When the disorder also couples adjacent nodes,    
strong coupling arguments are invoked\cite{Zirnbauer} which predict localization of 
quasiparticles and linearly vanishing density of states. Yet, even this case seems to 
suggest a scenario in which quasiparticle motion from ballistic turns directly into 
localized without crossing any intermediate diffusive regime.   

The standard non-linear $\sigma$-model approach applies the replica trick to average 
over disorder from the outset. The resulting fermion
interaction is then decoupled by introducing $Q$-matrix fields
in terms of which an effective action is derived after integrating out
the fermions. The saddle point solution is just the Born approximation, which   
provides, in the case of nodal Bogoliubov quasiparticles, a finite density of states 
hence a finite damping. Finally, long-wavelength transverse 
fluctuations are taken with respect to the saddle point leading  
to a non-linear $\sigma$-model action for the $Q$-matrix fields. 
However, unlike in conventional disordered systems, in this particular case an additional 
term may appear in the non-linear $\sigma$-model action, namely a so-called 
Wezz-Zumino-Novikov-Witten (WZW) term. It was actually   
argued\cite{Fendley,Fukui,Zirnbauer} that  
two opposite nodes share the same WZW term, while two adjacent ones have opposite WZW terms. 
As a consequence, when the two pairs of opposite nodes are uncoupled by disorder 
the WZW term is effective and the $\beta$-function of the spin-conductance flows towards 
an intermediate-coupling fixed point, signaling a delocalized behavior. 
On the contrary, when disorder couples all nodes together, 
the two WZW terms cancel exactly and the 
non-linear $\sigma$-model has no more protection against flowing towards 
a zero-conductance strong coupling regime characterized by a linearly vanishing density 
of states\cite{Fisher2}.

\bigskip
From the above discussion one might be lead to conclude that the 
agreement between the conclusions drawn with either methods is merely an accident  
which does not justify {\sl per se} the conventional non-linear $\sigma$-model approach when 
dealing with Dirac fermions. The main objection against the conventional non-linear 
$\sigma$-model is that it is not appropriate to 
start from a symmetry breaking saddle-point solution, associated to a mean-field-like finite 
density of states, when the outcome of including fluctuations is a vanishing density of states. 
Put in a different language, it is hard to believe in a method which starts by assuming 
a diffusive behavior if at the end it is discovered that a diffusive regime never appears.  
This criticism could invalidate also the results obtained with isotropic scattering, 
even though in this case the dimensionless coupling of the non-linear $\sigma$-model can be 
made small\cite{coupling} by assuming a large anisotropy of the 
Dirac dispersion (i.e. the velocities along and orthogonal to the Fermi surface).
A related objection that can be raised is that the non-linear $\sigma$-model approach 
to disordered systems is commonly believed to be valid for length scales longer than the 
mean free path and  
energy scales smaller than the inverse relaxation time $1/\tau$. On the contrary, 
the solution of the intra-node scattering problem demonstrates that disorder starts to 
affect for instance the density of states at energies of the order of the superconducting gap, 
hence much bigger than $1/\tau$, namely in the regime when quasiparticle 
motion should be still ballistic.

\medskip

It is the scope of the present work to clear up these inconsistencies 
of the non-linear $\sigma$-model approach to disordered $d$-wave supercondutors. We 
will demonstrate that the above, apparently contradicting,  
approaches can be actually reconciled. This is of particular interest since it 
provides further support to the  
standard non-linear $\sigma$-model approach based on the replica trick within the fermionic 
path-integral formalism, which remains so far one of the few available tools to 
deal with disorder in generic situations with a Fermi sea of interacting quasiparticles. 
Let us briefly summarize our main results.

First we are going to present a simple and straightforward way to 
extract the WZW term. Indeed it is well known  
how to derive the WZW term within field theories defined on a continuous space 
with Dirac-like dispersing particles.  
However in disordered lattice models the existence of such a term is not at all 
a common situation. We will show that the WZW term emerges quite naturally within 
the conventional derivation of the non-linear $\sigma$-model for disordered systems as 
a consequence of the non-analytical properties of the spectrum within the 
Brillouin zone. More specifically, the spin-current 
density in momentum space, $\mathbf{J}_{\mathbf{k}}$, in 
a $d$-wave superconductor is a $2\times 2$ matrix in the Nambu spinor space. 
The WZW term arises just because the vector product 
$\mathbf{J}_{\mathbf{k}}\wedge \mathbf{J}_{\mathbf{k}}$ is finite and actually 
gives a measure of the vorticity around each node.  
In the presence of purely intra-node impurity scattering we 
obtain the same WZW action of the non-abelian bosonization from the 1d mapping\cite{Nersesyan,Zirnbauer}, 
with however the inverse mean free path as a momentum upper cut-off instead of
the inverse lattice spacing which is usual the Ultra-Violet-regularizer of the 1d Dirac theory. 
In this context we elucidate the role of the WZW term in providing the
correct scaling behavior of the density of states depending on the impurity-potential range.

Another issue we clarify is the relationship between the coupling constant of 
the non-linear $\sigma$-model and the actual spin-conductance. 
The two quantities are known to coincide at the 
level of the Born approximation. However, rigorously speaking, the spin-conductance 
has to be calculated through a Kubo formula which involves advanced and retarded 
Green's functions. Since it makes a difference whether quasiparticles are right at 
the nodal points, with zero density of states, or slightly away from them, in which 
case the density of states is finite, one might wander whether the two quantities, 
spin-conductance and non-linear $\sigma$-model coupling, remain equal even beyond the 
Born approximation, especially in the case of intra-node and inter-opposite-node 
scattering. We will show that this is actually the case.

Finally we will show that the range of the impurity potential crucially affects the 
energy scale at which localization precursor effects starts to appear. In particular 
we will explicitly show that, keeping 
fixed the inverse relaxation time within the Born approximation, $1/\tau_0$, and 
increasing the relative weight of the long-range with respect to the short-range components of 
the disorder, leads to strong reduction below $1/\tau_0$ 
of the energy scale for the on-set of localization. This may provide a natural explanation 
to the partial lack of experimental evidences of quasiparticle localization in superconducting 
cuprates.

\medskip

The paper is organized as follows. In Section~II we introduce the model which we re-formulate 
within the fermionic path-integral formalism using the replica trick in Section~III. The global 
symmetries of the action are discussed in Section~IV. In Section~V we start the derivation 
of the non-linear $\sigma$-model which includes two terms. 
The ``conventional'' one is worked out in Section~VI, and its drawbacks discussed in 
Section~VII. The ``unconventional'' WZW term is derived in Section~VIII and its consequences 
for intra-nodal disorder are discussed in Section~IX. In Section~X 
we analyse the role of inter-nodal scattering processes and finally Section~XI is devoted to
the concluding remarks.

\section{The model}

The model we study is described by an Hartree-Fock Hamiltonian for $d$-wave superconductors 
in the presence of disorder   
\bea
{\cal{H}} &=& \sum_\ka \psi^\dagger_\ka \,\left[\epsilon_\ka\,\tau_3 
+ \Delta_\ka\, \tau_1 \right]\, \psi^{\phantom{\dagger}}_\ka 
+ \sum_{\ka\q} V(\q)\, \psi^\dagger_{\ka}\,\tau_3\,\psi^{\phantom{\dagger}}_{\ka+\q}
\nonumber \\
&\equiv&  \sum_\ka \psi^\dagger_\ka \,H^{(0)}_\ka\, \psi^{\phantom{\dagger}}_\ka 
+ \sum_{\ka\q} V(\q)\, \psi^\dagger_{\ka}\,\tau_3\,\psi^{\phantom{\dagger}}_{\ka+\q}.
\label{Ham}
\eea
Here $V(\q)$ is the impurity potential, 
$\psi^\dagger_\ka=(c^\dagger_{\ka\uparrow},\; c^\dagga_{-\ka\downarrow})$ 
the Nambu spinor,  $\Delta_\ka = \Delta\,(\cos k_xa - \cos k_y a)/2$ the 
superconducting gap with $d$-wave symmetry and $\epsilon_\ka$ the band-dispersion 
measured with respect to the chemical potential. 
The Pauli matrices $\tau_i$'s, $i=1,2,3$, act on the Nambu spinor components. 
The spectrum of the Bogolubov quasiparticles is given by 
\be
E_\ka = \sqrt{\epsilon_\ka^2 + \Delta_\ka^2},
\label{Ek}
\ee
and shows four nodal points at $\ka_1 = k_F\,(1,1)/\sqrt{2}$, $\ka_{\overline{1}}=-\ka_1$, 
$\ka_2=k_F\, (-1,1)/\sqrt{2}$ and $\ka_{\overline{2}}=-\ka_2$, 
$k_F$ being the Fermi momentum along the 
diagonals of the Brillouin zone. Around the nodes it is actually more convenient to rotate 
the axes by 45 degrees through 
\be
k_\xi = \frac{1}{\sqrt{2}}\left(k_x+k_y\right),\;\;
k_\eta = \frac{1}{\sqrt{2}}\left(k_y-k_x\right).
\label{xieta}
\ee
In the new reference frame, $\ka_1 = k_F\,(1,0)$ and $\ka_2=k_F\,(0,1)$. In what follows we 
define 
\be
\kap \equiv \ka - \ka_{a} 
\label{def:kap}
\ee
the momentum deviation from any of the four 
nodal points $\ka_a$, $a=1,\overline{1},2,\overline{2}$. For small $|\kap|\ll k_F$ the 
spectrum around nodes 1($\overline{1}$) and 2($\overline{2}$) is, respectively,  
\bea
E_{\ka_{1} + \kap} &\equiv& E_{1\, \kap} \simeq
\sqrt{v_F^2\, \kappa_\xi^2 + v_\Delta^2\,\kappa_\eta^2},
\label{E1}\\
E_{\ka_{2} + \kap} &\equiv& E_{2 \, \kap} \simeq  
\sqrt{v_F^2\, \kappa_\eta^2 + v_\Delta^2\,\kappa_\xi^2},
\label{E2}
\eea
thus having a conical Dirac-like form. Here $v_F = |\bnabla\epsilon_{\ka_a}|$ 
and $v_\Delta = |\bnabla\Delta_{\ka_a}|$.    

In the presence of disorder the motion of the gapless quasiparticles may become diffusive 
in the hydrodynamic regime. However diffusive propagators appear only in those channels 
which refer to conserved quantities. Hence, in superconductors, only thermal and spin density 
fluctuations might acquire a diffusive behavior. Let us therefore briefly discuss 
some properties of the spin current operator which are going to play an important role 
in our analysis.   

The $z$-component of the spin-current operator in the Nambu representation satisfies
\be
\q \cdot \sum_\ka  \psi^\dagger_\ka \,\J_{\ka}(\q)\, \psi^{\phantom{\dagger}}_{\ka+\q}
 = \sum_\ka  \psi^\dagger_\ka \,\left[
H^{(0)}_{\ka+\q}-H^{(0)}_\ka\right]\, \psi^{\phantom{\dagger}}_{\ka+\q},
\ee
hence, for $\q\to 0$,
\be
\J_{\ka}(\q) \rightarrow \J_{\ka}\equiv \bnabla\epsilon_\ka \,\tau_3 
+ \bnabla \Delta_\ka\, \tau_1.
\ee
Since the Pauli matrices anticommute, the following vector product turns out to be non-zero
\be
\J_{\ka}\wedge\J_{\ka} = 2 \,i\,\tau_2\,\bnabla\epsilon_\ka \wedge \bnabla \Delta_\ka. 
\label{JtimesJ} 
\ee
The vector product (\ref{JtimesJ}) actually 
probes the vorticity of the spectrum in momentum space; nodes ``1'' and ``$\overline{1}$'' 
have the same vorticity, $\J_{\ka}\wedge\J_{\ka} \to 2\, i\,\tau_2\,v_F\,v_\Delta$,
as well as nodes ``2'' and ``$\overline{2}$'', although opposite of node ``1'',
$\J_{\ka}\wedge\J_{\ka} \to -2\,i\,\tau_2\,v_F\,v_\Delta$. We are going to show that this 
property is crucial to uncover the physical behavior in the presence of 
disorder.  
 
\section{Path integral formulation}

To analyze the effect of disorder in this model, we are going to use a replica trick method 
within the fermionic path-integral approach. For that we associate to each fermionic 
operator Grassmann variables through 
\[
c^\dagger_{\ka\sigma}\rightarrow \overline{c}_{\ka\sigma},\;\;
c^\dagga_{\ka\sigma}\rightarrow c_{\ka\sigma}.
\]
Notice that, unlike the original fermionic operators, $\overline{c}_{\ka\sigma}$ 
and $c_{\ka\sigma}$ are independent variables. After introducing the Nambu spinors 
\[
\overline{\psi}_{n\ka}=\left(\overline{c}_{\ka\uparrow}(i\omega_n),\; 
c_{\ka\downarrow}(-i\omega_n)\right),
\;\;
\psi_{n\ka} = \left(
\begin{array}{c}
c_{\ka\uparrow}(i\omega_n)\\
\overline{c}_{\ka\downarrow}(-i\omega_n)\\
\end{array}
\right),
\]
where $\overline{c}_{\ka\sigma}(i\omega_n)$ and $c_{\ka\sigma}(i\omega_n)$ are the Grassmann 
variables in Matsubara frequencies, the path-integral action without disorder reads 
\be
{\cal{S}}_0 = \sum_n\,\sum_\ka \overline{\psi}_{n\ka}\, 
\left[\epsilon_\ka\,\tau_3 
+ \Delta_\ka\, \tau_1 -i\omega_n\right]\,\psi_{n\ka}.
\label{S0}
\ee
The disorder introduces the additional term
\be
{\cal{S}}_{imp} = \sum_n\,\sum_{\ka\q}\, V(\q)\, 
\overline{\psi}_{n\ka}\,\tau_3\,\psi_{n\ka+\q}.
\label{Simp}
\ee
Since $\overline{\psi}_{n\ka}$ and $\psi_{n\ka}$ are independent variables,  
the global transformation 
\be
\psi_{n\ka} \rightarrow {\rm e}^{i\theta\tau_2}\, \psi_{n\ka},\;\;
\overline{\psi}_{n\ka} \rightarrow \overline{\psi}_{n\ka}\, {\rm e}^{i\theta\tau_2},
\label{chiral-symm}
\ee
becomes allowed within the path-integral and is indeed 
a symmetry transformation of the full action ${\cal{S}}= {\cal{S}}_0+{\cal{S}}_{imp}$ 
when $\omega_n=0$. A finite frequency, $\omega_n\not=0$, spoils this symmetry. 
If, in addition, the disorder is long range, namely it does not induce inter-node scatterings, 
it is possible to define four independent symmetry transformations of the above kind, 
one for each node. 
This type of {\sl chiral} symmetry plays a crucial role for long range 
disorder, as was first emphasized in Ref.~[\onlinecite{Nersesyan}].

We notice that the disorder does not induce any mixing between different Matsubara frequencies, 
so that the action decouples into separate pieces, each one referring to a pair of 
opposite Matsubara frequencies, $\pm \omega_{n}$, which are coupled together 
by the superconducting term. For this reason we will just consider one of these pairs, 
with frequencies which we denote by $\pm \omega$, discarding  
all the others. This is enough to extract information about the 
quasiparticle density of states (DOS) at given frequency as well as about transport 
coefficients.   
  
\medskip

In order to derive a long-wavelength effective action, we resort to the replica-trick 
technique, hence we first 
introduce $N$ replicas of the Grassman variables, $\overline{c}_{\ka\sigma}(\pm i\omega)
\rightarrow \overline{c}_{a\ka\sigma}(\pm i\omega)$ and 
$c_{\ka\sigma}(\pm i\omega)
\rightarrow c_{a\ka\sigma}(\pm i\omega)$, with $a=1,\dots,N$. At the end we shall send 
$N\to 0$.  
Next we define the column vectors $c_\ka$ and $\overline{c}_\ka$, with $4N$ 
elements ($N$ replicas, 2 spins and 2 frequencies)
$c_{a\ka\sigma}(\pm i\omega)$ and $\overline{c}_{a\ka\sigma}(\pm i\omega)$, 
respectively. 
Finally we introduce new Nambu spinors through\cite{EL&K} 
\[
\Psi_{\ka} = \frac{1}{\sqrt{2}}
\left(
\begin{array}{c}
\overline{c}_{-\ka} \\
i\sigma_2 c_\ka \\
\end{array}
\right),
\]
as well as 
\[
\overline{\Psi}_\ka = \Psi_{-\ka}^t\, c^t,
\]
with the charge conjugacy operator defined by\cite{EL&K} 
\[
c = i\, \sigma_2\, \tau_1.
\]
Here the Pauli matrices $\sigma_1$, $\sigma_2$ and $\sigma_3$ act on the spin components  
of the column vectors $c_\ka$ and $\overline{c}_\ka$, while $\tau_1$, $\tau_2$ and $\tau_3$ 
act on the Nambu components (the diagonal elements refer to the particle-hole channels 
and the off-diagonal ones to the particle-particle channels). 
For later convenience, we also introduce the 
Pauli matrices $s_1$, $s_2$ and $s_3$,  
which act on the opposite frequency partners, as well as the identity matrices in 
the spin, Nambu and frequency subspaces, 
$\sigma_0$, $\tau_0$ and $s_0$, respectively. 

\medskip

By means of the above definitions,  the clean action can be written as 
\be
{\cal{S}}_0 = \sum_\ka \overline{\Psi}_{\ka}\, 
\left[\epsilon_\ka\,
+ i \, \Delta_\ka\, \tau_2\, s_1 -i\omega\, s_3\right]\,\Psi_{\ka}.
\label{S0-bis}
\ee
Since we are interested in the low-energy long-wavelength behavior, we 
shall focus our attention only in small areas around each of the four nodal points. 
In the rotated reference frame (\ref{xieta}), the nodes lie at 
$\ka_1 = k_F\, (1,0)$, $\ka_{\overline{1}}=-\ka_1$,
$\ka_2 = k_F\,(0,1)$ and $\ka_{\overline{2}}=-\ka_2$. Using the definition 
(\ref{def:kap}) for the small momentum deviations away from the nodal points,  
we introduce new Grassmann variables defined around each node through   
\[
c_{a,\kap}\equiv c_{\ka_a+\kap},\;\;
 \overline{c}_{a,\kap}\equiv \overline{c}_{\ka_a+\kap},
\]
$a=1,\overline{1},2,\overline{2}$, as well as the corresponding Nambu spinors
\be
\Psi_{1\kap}= 
\left(
\begin{array}{c}
\overline{c}_{\overline{1},-{\kap}} \\
i\sigma_2 c_{1,{\kap}} \\
\end{array}
\right),\,
\Psi_{\overline{1}{\kap}}= 
\left(
\begin{array}{c}
\overline{c}_{1,-{\kap}} \\
i\sigma_2 c_{\overline{1},{\kap}} \\
\end{array}
\right),
\Psi_{2\kap}= 
\left(
\begin{array}{c}
\overline{c}_{\overline{2},-{\kap}} \\
i\sigma_2 c_{2,{\kap}} \\
\end{array}
\right),
\Psi_{\overline{2}{\kap}}= 
\left(
\begin{array}{c}
\overline{c}_{2,-{\kap}} \\
i\sigma_2 c_{\overline{2},{\kap}} \\
\end{array}
\right).
\ee
One notices that for $a=1,2$  
\begin{eqnarray*}
\overline{\Psi}_{a{\kap}} &=& 
\frac{1}{\sqrt{2}}\left(-c_{\overline{a},{-\kap}}, 
-i\,\overline{c}_{a,{\kap}}\,\sigma_2\right) 
= \Psi_{\overline{a}-{\kap}}^t\, c^t,\\
\overline{\Psi}_{\overline{a}{\kap}} &=& 
\frac{1}{\sqrt{2}}\left(-c_{a,{-\kap}}, 
-i\,\overline{c}_{\overline{a},{\kap}}\,\sigma_2\right) 
= \Psi_{a-{\kap}}^t\, c^t.
\end{eqnarray*}
The non-disordered action expanded around the nodes reads 
\ba
{\cal{S}}_0 &=& \sum_{\kap}\,
\overline{\Psi}_{1{\kap}}\,\left[ v_F\, {\kappa}_\xi + 
i\,v_\Delta\, {\kappa}_\eta\, \tau_2 \, s_1\right]\, \Psi_{1{\kap}}
- \overline{\Psi}_{\overline{1}{\kap}}\,\left[ v_F\, {\kappa}_\xi + 
i\,v_\Delta\, {\kappa}_\eta\, \tau_2 \, s_1\right]\, \Psi_{\overline{1}{k}}
\\
&&~~ + \overline{\Psi}_{2{\kap}}\,\left[ v_F\, {\kappa}_\eta +
i\,v_\Delta\, {\kappa}_\xi\, \tau_2 \, s_1\right]\, \Psi_{2{\kap}}
- \overline{\Psi}_{\overline{2}{\kap}}\,\left[ v_F\, {\kappa}_\eta + 
i\,v_\Delta\, {\kappa}_\xi\, \tau_2 \, s_1\right]\, \Psi_{\overline{2}{k}}\\
&&~~ -i\omega\sum_{i=1,\overline{1},2,\overline{2}}\, 
\overline{\Psi}_{i{\kap}}\, s_3\,\Psi_{i{\kap}}. 
\ea

\medskip

We find useful to define new spinors with components in each of the nodes through  
\[
\Psi_{\kap}= 
\left(
\begin{array}{c}
\Psi_{1\kap} \\
\Psi_{2\kap} \\
\Psi_{\overline{1}\kap}\\
\Psi_{\overline{2}\kap}\\
\end{array}
\right),
\]
so that 
\[
\overline{\Psi}_{{\kap}} =  \left(\Psi^t_{\overline{1}-\kap},
\Psi^t_{\overline{2}-\kap},\Psi^t_{1-\kap},\Psi^t_{2-\kap}\right)\, c^t
= \Psi^t_{-\kap}\, \gamma_1\, c^t,
\]
where the Pauli matrices $\gamma$'s act on the ``$m$'' and ``$\overline{m}$'' 
subspace, namely connect two opposite nodes. 
This naturally leads to a new charge conjugacy defined through
\be
{\cal{C}} = c\,\gamma_1 = i\,\sigma_2\, \tau_1\, \gamma_1.
\label{new-C}
\ee
In what follows we denote as pair 1 the two opposite nodes 
``1'' and ``$\overline{1}$'', and as pair 2 the other two nodes, 
``2'' and ``$\overline{2}$''. Consequently we need 
to introduce also matrices in the two-pair subspace, 1 and 2, 
namely connecting adjacent nodes, which we will denote as 
$\rho_0$, the identity, $\rho_1$, $\rho_2$ and $\rho_3$, the three Pauli matrices. 

The clean action reads now   
\begin{eqnarray}
{\cal{S}}_0 &=& \sum_{\kap}\,\overline{\Psi}_{\kap}\, \gamma_3\, 
\frac{1}{2}\left(\kappa_\xi+\kappa_\eta\right)\, 
\left[ v_F + 
i\,v_\Delta\, \tau_2 \, s_1\right]\, \Psi_{\kap}
\nonumber\\
&&~~ + \overline{\Psi}_{\kap}\, \gamma_3\,\rho_3\,  
\frac{1}{2}\left(\kappa_\xi-\kappa_\eta\right)
\left[ v_F - 
i\,v_\Delta\, \tau_2 \, s_1\right]\, \Psi_{\kap}
\; - \; i\omega\,\sum_{\kap}\,\overline{\Psi}_{\kap}\, s_3\, \Psi_{\kap}\\
&&~~\equiv \sum_{\kap} \overline{\Psi}_{\kap}\,\left[ {\cal{H}}_0(\kap) 
-i\omega\,s_3\right]\, \Psi_{\kap}
\label{S0-tris}
\end{eqnarray}
We notice that within this path-integral formulation, the four 
independent chiral 
symmetry transformations of the form (\ref{chiral-symm}) translate into 
\[
\Psi \rightarrow T\, \Psi,\;\; \overline{\Psi}\rightarrow \overline{\Psi}\, 
{\cal{C}}\,T^t\,{\cal{C}}^t,
\]
with $T$ given by any of the following expressions  
\bea
T &=& {\rm e}^{i\,\theta\,\tau_2\, s_1\,\gamma_0\, \rho_0\, \bm{\sigma}\cdot \bm{u}},
\label{chiral-00T}\\
T &=& {\rm e}^{i\,\theta\,\tau_2\, s_1\,\gamma_3 \, \rho_0 },
\label{chiral-30S}\\
T &=& {\rm e}^{i\,\theta\,\tau_2\, s_1\,\gamma_0 \, \rho_3\, \bm{\sigma}\cdot \bm{u}},
\label{chiral-03T}\\
T &=& {\rm e}^{i\,\theta\,\tau_2\, s_1\,\gamma_3 \, \rho_3 },
\label{chiral-33S}
\eea
where $\theta$ is a phase factor and $\bm{u}$ an arbitrary unit vector.

We assume that the scattering potential $V(\q)=V(-\q)^*$ 
induced by disorder has independent components 
which act inside each node, $V(\q)\equiv U_0(\q)$, 
between opposite nodes, $V(\q+2\ka_1)\equiv U_1(\q)$ and $V(\q+2\ka_2)\equiv U_2(\q)$,  
and between adjacent nodes, $V(\q+\ka_1-\ka_2)\equiv U_{12}(\q)$ and 
$V(\q+\ka_1+\ka_2)\equiv U_{1\overline{2}}(\q)$, so that the impurity contribution to 
the action is
\bea
{\cal{S}}_{imp} &=& \sum_{\kap,\q}\,U_0(\q)\, \overline{\Psi}_{\kap}\, 
\Psi_{\kap+\q} \nonumber \\
&& + \sum_{\kap,\q}\,U_1(\q)\; \overline{\Psi}_{1\kap}\,
\Psi_{\overline{1}\kap+\q} 
+ U_1(\q)^*\; \overline{\Psi}_{\overline{1}\kap+\q}\, \Psi_{1\kap} \nonumber \\
&& + \sum_{\kap,\q}\,U_2(\q)\; \overline{\Psi}_{2\kap}\, \Psi_{\overline{2}\kap+\q} 
+ U_2(\q)^*\; \overline{\Psi}_{\overline{2}\kap+\q}\,
\Psi_{2\kap} \nonumber \\
&& + \sum_{\kap,\q}\,U_{12}(\q)\; \left[\overline{\Psi}_{1\kap}\,
\Psi_{2\kap+\q} 
+ \overline{\Psi}_{\overline{2}\kap}\,
\Psi_{\overline{1}\kap+\q} \right] 
+ U_{12}(\q)^*\; \left[\overline{\Psi}_{2\kap+\q}\,
\Psi_{1\kap} 
+ \overline{\Psi}_{\overline{1}\kap+\q}\,
\Psi_{\overline{2}\kap} \right] \nonumber \\ 
&& + \sum_{\kap,\q}\,U_{1\overline{2}}(\q)\; \left[\overline{\Psi}_{1\kap}\,
\Psi_{\overline{2}\kap+\q} +\overline{\Psi}_{2\kap}\,
\Psi_{\overline{1}\kap+\q}\right] + 
U_{1\overline{2}}(\q)^*\; \left[\overline{\Psi}_{\overline{2}\kap+\q}\,\Psi_{1\kap} 
+ \overline{\Psi}_{\overline{1}\kap+\q}\,\Psi_{2\kap}\right].\nonumber \\
\label{S-disorder-1}
\eea
Apart from the first term in the right hand side of (\ref{S-disorder-1}), 
we have for convenience kept the node labels. Since the following relations hold:
\ba
&&\sum_\kap\, \overline{\Psi}_{\overline{2}\kap}\,
\Psi_{\overline{1}\kap+\q} = 
\sum_\kap\, \Psi_{2 - \kap}^t\,c^t\,c^t\,
\overline{\Psi}_{1 -\kap-\q}^t 
= \sum_\kap\, \overline{\Psi}_{1 \kap}\,
{\Psi}_{2 \kap+\q},\\
&&\sum_\kap\, \overline{\Psi}_{2\kap}\,
\Psi_{\overline{1}\kap+\q} = 
\sum_\kap\, \Psi_{\overline{2} - \kap}^t\,c^t\,c^t\,
\overline{\Psi}_{1 -\kap-\q}^t =
 \sum_\kap\, \overline{\Psi}_{1 \kap}\,
{\Psi}_{\overline{2} \kap+\q}.
\ea
we can rewrite (\ref{S-disorder-1}) in the following way:
\bea
{\cal{S}}_{imp} &=& \sum_{\kap,\q}\,U_0(\q)\, \overline{\Psi}_{\kap}\, 
\Psi_{\kap+\q} \nonumber \\
&& + \sum_{\kap,\q}\,U_1(\q)\; \overline{\Psi}_{1\kap}\,
\Psi_{\overline{1}\kap+\q} 
+ U_1(\q)^*\; \overline{\Psi}_{\overline{1}\kap+\q}\, \Psi_{1\kap} \nonumber \\
&& + \sum_{\kap,\q}\,U_2(\q)\; \overline{\Psi}_{2\kap}\, \Psi_{\overline{2}\kap+\q} 
+ U_2(\q)^*\; \overline{\Psi}_{\overline{2}\kap+\q}\,
\Psi_{2\kap} \nonumber \\
&& + 2\, \sum_{\kap,\q}\,U_{12}(\q)\; \overline{\Psi}_{1\kap}\,
\Psi_{2\kap+\q} 
+ U_{12}(\q)^*\; \overline{\Psi}_{2\kap+\q}\,
\Psi_{1\kap} \nonumber \\ 
&& + 2\, \sum_{\kap,\q}\,U_{1\overline{2}}(\q)\; \overline{\Psi}_{1\kap}\,
\Psi_{\overline{2}\kap+\q} + 
U_{1\overline{2}}(\q)^*\; \overline{\Psi}_{\overline{2}\kap+\q}\,\Psi_{1\kap}. 
\label{S-disorder}
\eea

Finally we go back in real space, which now corresponds to the low-energy continuum limit 
of the original lattice model, and obtain the clean action 
\bea
{\cal{S}}_0 &=& \int d\rr\, \overline{\Psi}(\rr)\, \frac{1}{2}\,\gamma_3\,  
\left( -i\partial_\xi -i\partial_\eta\right)\, 
\left[v_F + i\,v_\Delta\,\tau_2\,s_1\right]\, \Psi(\rr) \nonumber \\
&& + \int d\rr\, \overline{\Psi}(\rr)\, \frac{1}{2}\,\gamma_3\, \rho_3\,  
\left( -i\partial_\xi + i\partial_\eta\right)\, 
\left[v_F - i\,v_\Delta\,\tau_2\,s_1\right]\, \Psi(\rr) \nonumber \\ 
&& -i\omega\, \int d\rr\, \overline{\Psi}(\rr)\,s_3\, \Psi(\rr)
\equiv \int d\rr\, \overline{\Psi}(\rr)\,\left[ {\cal{H}}_0(\rr) -i\omega\,s_3\right]\, 
\Psi(\rr),\label{S0-realspace}
\eea 
and the impurity term 
\bea
{\cal{S}}_{imp} &=& \int d\rr\, U_0(\rr)\; \overline{\Psi}(\rr)\, \Psi(\rr)\nonumber \\
&& + U_1(\rr)\;\overline{\Psi}_1(\rr)\,\Psi_{\overline{1}}(\rr) 
+ U_1(\rr)^*\;\overline{\Psi}_{\overline{1}}(\rr)\,
\Psi_1(\rr) \nonumber \\
&& + U_2(\rr)\;\overline{\Psi}_2(\rr)\,
\Psi_{\overline{2}}(\rr) 
+ U_2(\rr)^*\;\overline{\Psi}_{\overline{2}}(\rr)\,\Psi_2(\rr) \nonumber \\
&& + 2\, U_{12}(\rr)\;\overline{\Psi}_1(\rr)\,\Psi_2(\rr) 
+ 2\, U_{12}(\rr)^*\;\overline{\Psi}_2(\rr)\,\Psi_1(\rr) \nonumber \\
&& + 2\, U_{1\overline{2}}(\rr)\;\overline{\Psi}_1(\rr)\,
\Psi_{\overline{2}}(\rr) 
+ 2\, U_{1\overline{2}}(\rr)^*\;\overline{\Psi}_{\overline{2}}(\rr)\,
\Psi_1(\rr) .\label{Simp-realspace}
\eea
We further assume that the disorder is $\delta$-like correlated with  
\be
\begin{array}{ll}
\langle U_0(\rr)\rangle = 0, & \langle U_0(\rr)\,U_0(\rr^\prime)\rangle = 
u^2\, \delta\left(\rr-\rr^\prime\right),\\
\langle U_1(\rr)\rangle = 0, & \langle U_1(\rr)\,U_1(\rr^\prime)^*\rangle = 
2\, v^2\, \delta\left(\rr-\rr^\prime\right),\\
\langle U_2(\rr)\rangle = 0,& \langle U_2(\rr)\,U_2(\rr^\prime)^*\rangle = 
2\,v^2\, \delta\left(\rr-\rr^\prime\right),\\
\langle U_{12}(\rr)\rangle = 0, & \langle U_{12}(\rr)\,U_{12}(\rr^\prime)^*\rangle = 
2\,w^2\, \delta\left(\rr-\rr^\prime\right),\\
\langle U_{1\overline{2}}(\rr)\rangle = 0,&
\langle U_{1\overline{2}}(\rr)\,U_{1\overline{2}}(\rr^\prime)^*\rangle = 
2\,w^2\, \delta\left(\rr-\rr^\prime\right).\\
\end{array}
\label{disorder-P}
\ee

\medskip

We conclude by noticing that ({\sl i}) if only $u\not = 0$ 
all transformations (\ref{chiral-00T})--(\ref{chiral-33S}) leave the action invariant; 
({\sl ii}) if $u\not = 0$ and $v\not = 0$ only (\ref{chiral-00T}) and (\ref{chiral-03T}) 
are allowed; ({\sl iii}) finally if $u\not = 0$, $v\not = 0$ and $w\not = 0$ only 
(\ref{chiral-00T}) remains.

\section{Symmetry properties} 

Since we have been obliged to introduce so many Pauli matrices, including the identity 
denoted as the zeroth Pauli matrix, for sake of clarity 
we prefer to start this Section by first summarizing in Table~\ref{pauli-matrices} the 
subspaces in which any of them act.
\begin{table}[t]
\centerline{\begin{tabular}{||c|c||}\hline\hline
Pauli matrices & subspace of action \\ \hline
$\sigma_0,\sigma_1,\sigma_2,\sigma_3$ & spin components\\ \hline 
$\tau_0,\tau_1,\tau_2,\tau_3$ & Nambu components \\ \hline 
$s_0,s_1,s_2,s_3$ & opposite frequency partners \\ \hline 
$\gamma_0,\gamma_1,\gamma_2,\gamma_3$ & opposite nodal points \\ \hline
$\rho_0,\rho_1,\rho_2,\rho_3$ & the two pairs of opposite nodes \\ \hline\hline
\end{tabular}}
\caption{The various two-components subspaces in which each set of Pauli matrices act}
\label{pauli-matrices}
\end{table}

Now, let us uncover all global symmetry transformations. We start assuming 
that only intra-node disorder is present, namely $U_0\not=0$ but 
$U_1=U_2=U_{12}=U_{1\overline{2}}=0$. 

If the frequency is zero, the action $S_0+S_{imp}$ is invariant under 
unitary global transformations $T$ such that 
\begin{eqnarray*}
{\cal{C}}^t \, T^t\,{\cal{C}} \, \gamma_3\, T &=& \gamma_3, \\
{\cal{C}}^t \, T^t\,{\cal{C}} \, \gamma_3\, \tau_2\,s_1\, T 
&=& \gamma_3\, \tau_2 \,s_1, \\
{\cal{C}}^t \, T^t\,{\cal{C}} \, \gamma_3\,\rho_3\, T &=& \gamma_3\,\rho_3, \\
{\cal{C}}^t \, T^t\,{\cal{C}} \, \gamma_3\, \rho_3\, \tau_2\,s_1\, T 
&=& \gamma_3\, \rho_3\,\tau_2 \,s_1, \\
{\cal{C}}^t \, T^t\,{\cal{C}} \, T &=& 1.
\end{eqnarray*}
They imply that 
\be
\label{cTc}
{\cal{C}}^t \, T^t\,{\cal{C}} = T^{-1},
\ee
as well as that 
\be
\label{three[]}
\left[T,\gamma_3\right]=0,\;\;
\left[T,\rho_3\right]=0,\;\;
\left[T,\tau_2\,s_1\right]=0.
\ee
We parametrize $T$ as
\be
T = \frac{1}{2}\left(T_1+T_2\right)\, \rho_0 
+ \frac{1}{2}\left(T_1-T_2\right)\, \rho_3,
\label{parametrization-T}
\ee
where the suffix ``1'' refers to the pair 1 
(opposite nodes 1 and $\overline{1}$)   
and ``2'' to the pair 2 (opposite nodes 2 and $\overline{2}$). 
The symmetry modes for pair 1 and 2 can be   
parametrized by the  $16N\times 16N$ unitary operators
\[
T_{1(2)} = \frac{1}{2}\left(\gamma_0+\gamma_3\right)\, V_{1(2)} 
+ \frac{1}{2}\left(\gamma_0-\gamma_3\right)\,  
V_{\overline{1}(\overline{2})} \equiv
\left(
\begin{array}{cc}
V_{1(2)} & 0 \\
0 & V_{\overline{1}(\overline{2})} \\
\end{array}
\right),
\]
where the $V$'s are $8N\times 8N$ unitary matrices. 
Because of (\ref{cTc}) we get 
\[
{\cal{C}}^t\,T_{1(2)}^t\,{\cal{C}} = T_{1(2)}^{-1} = T_{1(2)}^\dagger,
\]
and $V_{1(2)}$ and $V_{\overline{1}(\overline{2})}$ are 
actually not independent as 
\be
\label{notindependent}
c^t \, V_{1(2)}^t \, c = V_{\overline{1}(\overline{2})}^\dagger.
\ee
Therefore
\[
T_{1(2)} =\left(
\begin{array}{cc}
V_{1(2)} & 0 \\
0 & c^t \, \left(V_{1(2)}^\dagger\right)^t \, c \\
\end{array}
\right),
\]
is indeed parametrized by a single  $8N\times 8N$ unitary matrix $V_{1(2)}$. 
Since the two pairs of nodes behave similarly, 
in what follows we drop the suffices 1 and 2. 
According to (\ref{three[]}) we still need to impose that 
\be
\left[V,\tau_2\,s_1\right] = 0.
\label{V-cond-symm}
\ee
To this purpose let us introduce the following unitary operation 
\be
U = \left[\frac{1}{2}\left(\tau_0+\tau_2\right) - 
\frac{1}{2}\left(\tau_0-\tau_2\right)\,s_3\right]
\, {\rm e}^{-i\frac{\pi}{4}\,s_2}.
\label{U-unitary}
\ee
which transforms
\[
\Psi\rightarrow U\, \Psi,\;\;
\overline{\Psi}\rightarrow \Psi^t \, U^t\, c^t = \Psi^t\, c^t\, U^\dagger = 
\overline{\Psi}\, U^\dagger,
\] 
so that for any operator $\cal{O}$   
\[
\overline{\Psi}\,{\cal{O}}\, \Psi \rightarrow  \overline{\Psi}\,U^\dagger\, {\cal{O}}\, 
U\, \Psi.
\]

One readily shows that 
\be
U^\dagger\, \tau_2\, s_1 \, U = s_3,\;\; 
U^\dagger\, s_3\, U = -s_1,
\label{U-c}
\ee
hence (\ref{V-cond-symm}) transforms into 
\be
\left[V,s_3\right] = 0,
\label{V2-cond-symm}
\ee
which is fulfilled by the general expression  
\be
\label{eqV}
V = \frac{1}{2}\left(A+B\right)\, s_0 + 
\frac{1}{2}\left(A-B\right)\, s_3, 
\ee
with $A$ and $B$ being independent $4N\times 4N$ unitary matrices. Therefore 
the original symmetry turns out to be U$(4N)\times$U$(4N)$ for pair 1, and analogously 
for pair 2, in total U$(4N)\times$U$(4N)\times$U$(4N)\times$U$(4N)$.

We notice that, 
in the presence of a finite frequency, $\omega\not= 0$, we shall further impose 
through (\ref{U-c}) that 
\[
\left[V,s_1\right]=0,
\]
which is satisfied by  
\[
V= C\, s_0,
\]
with $C$ belonging to U$(4N)$. Therefore the symmetry is lowered by the frequency down  
to U$(4N)$ for the pair 1 times U$(4N)$ for the pair 2. 

Within the non-linear $\sigma$-model approach to disordered systems, the frequency 
plays actually the role of a symmetry-breaking field. This leads to the identification 
of the transverse modes $V^\perp$ as those satisfying  
\[
s_1\, V^\perp \, s_1 = \left(V^\perp\right)^{-1} = 
\left(V^\perp\right)^\dagger,
\]
implying $A=B^\dagger$ in Eq.(\ref{eqV}) so that we can write
\be
V^\perp = \frac{1}{2}\left[\sqrt{G} + \sqrt{G^\dagger}\right]\, s_0 
+ \frac{1}{2}\left[\sqrt{G} - \sqrt{G^\dagger}\right]\, 
s_3,\label{transverse}
\ee
where $G$ belongs to U$(4N)$. In other words the coset 
space spanned by the transverse modes is still a group, namely U$(4N)$. It is 
convenient to factorize out of $G$ the abelian component. For that we 
write
\be
G = {\rm e}^{i\sqrt{\frac{\pi}{N}}\,\phi}\; g,
\label{Gvsg}
\ee
where $\phi$ is a scalar and $g$ belongs to SU$(4N)$.
The forms (\ref{transverse}) and (\ref{Gvsg}) will be useful in the following
to express the non-linear $\sigma$-model directly in terms of g and $\phi$.
After the transformation (\ref{U-unitary}) from Eq.(\ref{notindependent}) 
we find that  
\be
V^\perp_{\overline{1}(\overline{2})} = 
\left[
\tau_1\, \sigma_2\,\left(V^\perp_{1(2)}\right)^\dagger \, 
\tau_1\, \sigma_2\right]^t 
= \left[
\tau_1\, \sigma_2\,s_1\, V^\perp_{1(2)}\,s_1\,  
\tau_1\, \sigma_2\right]^t,
\label{Vbar2}
\ee
which leads to  
\be
\phi_{\overline{1}(\overline{2})} = - \phi_{1(2)},\;\;
g_{\overline{1}(\overline{2})} = \tau_1\,\sigma_2\, \left(g_{1(2)}^\dagger\right)^t\, 
\sigma_2\,\tau_1.
\label{gbar-g}
\ee
\medskip

Let us conclude by showing what would change for more general disorder potential. 
First let us consider the case in which the disorder also contains terms which 
couple opposite nodes, 
{\sl i.e.} $U_1\not = 0$ and $U_2\not = 0$. In this case the $\gamma_3$-modes 
are not anymore allowed by symmetry, so we have to impose that 
$V_{1(2)}=V_{\overline{1}(\overline{2})}$, namely, through (\ref{gbar-g}), that 
$\phi_{1(2)}=0$ and  
\be
g_{1(2)}^\dagger = g_{1(2)}^{-1} = \tau_1\,\sigma_2\, g_{1(2)}^t\, \sigma_2\, \tau_1. 
\label{Sp(2N)}
\ee
The coset becomes now the group Sp$(2N)$ of unitary-symplectic 
matrices (also called USp$(4N)$) for pair 1 and analogously for pair 2, 
{\sl i.e.} Sp$(2N)\times$Sp$(2N)$. 

\medskip

Finally, if also scattering between adjacent nodes is allowed, $U_{12}\not=0$ 
and $U_{1\overline{2}}\not =0$, then also the $\rho_3$ modes do not leave the action invariant. 
In this case we have to impose that $V^\perp_1 = V^\perp_2$, so 
that the coset space is the group Sp$(2N)$.

\section{The non-linear $\sigma$-model}      
In what follows we begin analyzing just the case in which the disorder only induces 
intra-node scattering processes. At the end we will return back to the most general case.
Within the replica trick, we can average the action over the disorder probability 
distribution, after which (\ref{Simp-realspace}) with $v=w=0$ [see Eq.~(\ref{disorder-P})] 
transforms into 
\be
{\cal{S}}_{imp} = - \frac{u^2}{2}\,\int d\rr\, 
\left[\overline{\Psi}(\rr)\Psi(\rr)\right]^2.
\label{NLSM-1}
\ee
We define $32N\times 32N$ matrices $X(\rr)$ by 
\[
X(\rr)\equiv \Psi(\rr)\,\overline{\Psi}(\rr),
\]
so that (\ref{NLSM-1}) can be also written as 
\be
{\cal{S}}_{imp} =  \frac{u^2}{2}\,\int d\rr\, 
{\rm Tr}\left( X(\rr)\, X(\rr)\right).
\label{NLSM-2}
\ee
By means of an Hubbard-Stratonovich transformation one can show that 
\be
\exp\left\{-\frac{u^2}{2}\,\int d\rr\, 
{\rm Tr}\left( X(\rr)\, X(\rr)\right) \right\} 
= \int\, {\cal{D}}Q(\rr)\, 
\exp\left\{ -\int d\rr\; \frac{1}{2u^2}{\rm Tr}\left[Q(\rr)^2\right] 
- i\,{\rm Tr}\left[Q(\rr)\, X(\rr)\right]\right\},\label{NLSM-3}
\ee
with $Q(\rr)$ being hermitian $32N\times 32N$ auxiliary matrix fields. 

In conclusion the full action, (\ref{S0-realspace}) plus (\ref{NLSM-3}), becomes 
\bea
{\cal{S}} &=& \frac{1}{2u^2}\int d\rr\,{\rm Tr}\left[Q(\rr)^2\right]\nonumber \\
&& + \int d\rr\, \overline{\Psi}(\rr)\,\left[
{\cal{H}}_0(\rr) - i\,Q(\rr) - i\omega\,s_3\right]\,\Psi(\rr).\label{NLSM-4}
\eea   
One obtains the effective action which describes the auxiliary field $Q(\rr)$ by integrating 
out the Grassmann variables, thus getting
\be
{\cal{S}}\left[Q\right] = \frac{1}{2u^2}\int d\rr\,{\rm Tr}\left[Q(\rr)^2\right] 
\;-\; \frac{1}{2}{\rm Tr}\,\ln\, \left[iQ+i\omega\,s_3 - {\cal{H}}_0\right].
\ee

\medskip

We further proceed in the derivation of a long-wavelength effective action for 
$Q(\rr)$ by following the conventional approach. First of all we  calculate the saddle point 
expression of $Q(\rr)$, which we denote by $Q_0\, s_3$, assuming it is uniform,  
in the presence of an infinitesimal symmetry breaking term $\omega\,s_3$. Then we  
neglect longitudinal fluctuations and parametrize the actual $Q(\rr)$ by 
\be
Q(\rr) \simeq T(\rr)^{-1}\, \,Q_0\, s_3\, T(\rr) = 
\frac{1}{2}\left(\rho_0+\rho_3\right)\, Q_1(\rr) + 
\frac{1}{2}\left(\rho_0-\rho_3\right)\, Q_2(\rr), 
\label{NLSM-5}
\ee
where through (\ref{parametrization-T}) $Q_{ab}(\rr)=\delta_{ab}\, Q_{a}(\rr)$ with 
\be
Q_{a}(\rr) = T_a(\rr)^{-1}\, \,Q_0\, s_3\, T_a(\rr) 
= Q_0\, s_3\, T_a(\rr)^2 = T_a(\rr)^{-2}\, \,Q_0\, s_3,
\label{Qab}
\ee
being the auxiliary field in the pair subspace $a,b=1,2$. Notice that, even though the 
most general $Q(\rr)$ would couple the nodes together, 
namely would include off-diagonal elements $Q_{12}(\rr)$, (\ref{NLSM-5}) does not 
contain any mixing term. This simply reflects that the off-diagonal components 
are not diffusive, hence massive. 

In order to derive the long-wavelength action we 
find it convenient to decompose the action into the real part 
\bea
{\cal{S}}_{NL\sigma M} &=& \frac{1}{2u^2}\int d\rr\,{\rm Tr}\left[Q(\rr)^2\right] 
\nonumber \\
&& - \frac{1}{4}{\rm Tr}\,\ln\,\left[iQ+i\omega\,s_3 - {\cal{H}}_0\right] 
- \frac{1}{4}{\rm Tr}\,\ln\, \left[-iQ-i\omega\,s_3 - {\cal{H}}_0^\dagger\right],
\label{S-real}
\eea
which, as we shall see, gives the conventional non-linear $\sigma$-model, 
and the imaginary part
\be
{\cal{S}}_{\Gamma} = - \frac{1}{4}{\rm Tr}\,\ln\,\left[iQ+i\omega\,s_3 - {\cal{H}}_0\right] 
+ \frac{1}{4}{\rm Tr}\,\ln\,\left[-iQ-i\omega\,s_3 - {\cal{H}}_0^\dagger\right],
\label{S-imaginary}
\ee
which we will show gives rise to a WZW term.    

\section{Conventional $\sigma$-model}
By means of (\ref{NLSM-5}), the real part of the action, (\ref{S-real}), 
can be written as 
\be
{\cal{S}}_{NL\sigma M} = \frac{{\rm V_{eff}}}{2u^2}\, 32N\, Q_0^2  
- \frac{1}{4}{\rm Tr}\,\ln\left[{\cal{G}}^{-1}\right],
\ee  
where $V_{eff}$ is the effective volume corresponding to the long-wavelength theory (roughly 
speaking $V_{eff}=V/4$, since we have implicitly folded the Brillouin zone into a single 
quadrant, in order to make all nodes coincide), and  
\bea
{\cal{G}}^{-1} &=& {\cal{H}}_0\,{\cal{H}}_0^\dagger + Q_0^2 + \omega^2 
+ \omega\left(Q\,s_3 + s_3\,Q\right) + i {\cal{H}}_0\, Q - i Q\,{\cal{H}}_0^\dagger
\nonumber \\
&\equiv& {\cal{G}}_0^{-1} - \Sigma,
\label{NLSM-6}
\eea
where
\ba
\mathcal{G}_0^{-1} &=& {\cal{H}}_0\,{\cal{H}}_0^\dagger + Q_0^2 + \omega^2,\\
\Sigma &=& - \omega\left\{Q,s_3\right\} - i {\cal{H}}_0\, Q + i Q\,{\cal{H}}_0^\dagger.
\ea
We notice that 
$
s_3\,{\cal{H}}_0\,s_3 = {\cal{H}}_0^\dagger,
$
therefore 
\[
i\, {\cal{H}}_0\, Q - i\, Q\,{\cal{H}}_0^\dagger = 
i\, {\cal{H}}_0\, Q\, s_3\, s_3 - i\,Q\,s_3\,{\cal{H}}_0\,s_3 = 
i\, \left[{\cal{H}}_0,Q\, s_3\right]\, s_3 = 
\J \cdot \bnabla Q,
\]
where we made use of the equivalence $Q\, s_3 \equiv T^{-2}$ and 
of the expression of the spin current operator in the long wave-length limit
\be
\J(\rr) = i\, \left[{\cal{H}}_0(\rr),\rr\right].
\label{def-current}
\ee
Analogously one can show that 
\[
i\, {\cal{H}}_0\, Q - i\, Q\,{\cal{H}}_0^\dagger = 
i\, s_3\,{\cal{H}}_0^\dagger\, s_3\, Q - i\,s_3\,s_3\,Q\,{\cal{H}}_0^\dagger = 
 - i\, s_3\, \left[s_3\, Q,{\cal{H}}_0^\dagger\right] = 
\bnabla Q \cdot \J^\dagger,
\]
so that the self-energy operator can be written as 
\be
\Sigma = 
- \omega\left\{Q,s_3\right\} - \J\cdot \bnabla Q 
= 
- \omega\left\{Q,s_3\right\} - \bnabla Q\cdot \J^\dagger.
\label{Sigma}
\ee

We then expand the action up to second order in $\Sigma$ and obtain  
\bea
{\cal{S}}_{NL\sigma M}&=& \frac{{\rm V_{eff}}}{2u^2}\, 32N\, Q_0^2 
- \frac{1}{4}{\rm Tr}\,\ln\left[{\cal{G}}^{-1}\right] \nonumber \\
&=& \frac{{\rm V_{eff}}}{2u^2}\, 32N\, Q_0^2 +  
\frac{1}{4}{\rm Tr}\,\ln \left[{\cal{G}}_0\right] 
- \frac{1}{4}{\rm Tr}\,\ln\left[1 - {\cal{G}}_0\,\Sigma\right] \nonumber \\
&=& \frac{{\rm V_{eff}}}{2u^2}\, 32N\, Q_0^2 + 
\frac{1}{4}{\rm Tr}\,\ln\left[{\cal{G}}_0\right]   
+ \frac{1}{4}{\rm Tr}\left[{\cal{G}}_0\,\Sigma\right] + 
\frac{1}{8}{\rm Tr}\left[{\cal{G}}_0\,\Sigma\,{\cal{G}}_0\,\Sigma\right]\nonumber \\
&=& \frac{{\rm V_{eff}}}{2u^2}\, 32N\, Q_0^2+\frac{1}{4}{\rm Tr}\,\ln\left[{\cal{G}}_0\right] 
- \frac{\omega}{2}{\rm Tr}\left[Q\, s_3\,{\cal{G}}_0\right]\nonumber \\
&& + \frac{\omega^2}{8}{\rm Tr}\left[
\mathcal{G}_0\,\left\{Q,s_3\right\}\,\mathcal{G}_0\,\left\{Q,s_3\right\}\right]
+ \frac{1}{8}{\rm Tr}\left[{\cal{G}}_0\,\vec{\nabla}Q\cdot \J^\dagger\, 
{\cal{G}}_0\, \J\cdot \vec{\nabla}Q\right].
\label{NLSM-7}
\eea  

Notice that we have arrived to  ${\cal{S}}_{NL\sigma M}$
in terms of $\vec{\nabla}Q$ without passing through a gauge transformation to carry out the 
gradient expansion so avoiding any problem related to a proper treatment of the Jacobian
determinant\cite{Zirnbauer}.

\subsection{Saddle point equation}

In momentum space one finds that 
\ba
{\cal{H}}_{0\kap}{\cal{H}}_{0\kap}^\dagger &=& 
\frac{1}{2}\left(\kappa_\xi^2+\kappa_\eta^2\right)\left(v_F^2+v_\Delta^2\right) 
+\rho_3\, \frac{1}{2}\left(\kappa_\xi^2-\kappa_\eta^2\right)\left(v_F^2-v_\Delta^2\right)\\
&=& \frac{1}{2}\left(E_{1\kap}^2+E_{2\kap}^2\right)
+\rho_3\, \frac{1}{2}\left(E_{1\kap}^2-E_{2\kap}^2\right),
\ea
where $E_{1\kap}$ is the spectrum of quasiparticles around nodes ``1'' and ``$\overline{1}$'', 
and analogously for $E_{2\kap}$, see (\ref{E1}) and (\ref{E2}). 
Therefore
\[
 {\cal{G}}_{0\kap} = \frac{1}{2}\left[{\cal{G}}_{1\kap} + {\cal{G}}_{2\kap}\right]
+ \rho_3\, \frac{1}{2}\left[{\cal{G}}_{1\kap} - {\cal{G}}_{2\kap}\right],
\]
where 
\[
{\cal{G}}_{1\kap} = \frac{1}{E_{1\kap}^2 + Q_0^2+\omega^2},
\qquad
{\cal{G}}_{2\kap} = \frac{1}{E_{2\kap}^2 + Q_0^2+\omega^2},
\]
so that 
\be
{\cal{G}}_{0\kap}^2 = \frac{1}{2}\left[
{\cal{G}}_{1\kap}^2 + {\cal{G}}_{2\kap}^2\right]
+ \rho_3\, \frac{1}{2}\left[{\cal{G}}_{1\kap}^2 - {\cal{G}}_{2\kap}^2\right]
\label{GG}
\ee

The saddle-point equation is obtained by taking $Q(\mathbf{r})=Q_0$ and 
minimizing the action, which leads to the self-consistency equation 
\be
\frac{Q_0}{Q_0+\omega} = u^2\,\frac{1}{V}\, \sum_\ka \frac{\displaystyle 1}
{\displaystyle E_\ka^2 + \left(Q_0+\omega\right)^2}, 
\label{saddle-point}
\ee
originally derived in Ref.~[\onlinecite{Lee}]. The solution of this equation, 
in the limit $\omega \rightarrow 0$, reads
\[
Q_0 = \Lambda \, \exp\left(-\frac{\pi v_F v_\Delta}{2u^2}\right),
\]
where $\Lambda$ is an ultraviolet cut-off which is roughly the energy scale above 
which the spectrum deviates appreciably from a linear one, in other words 
$\Lambda \simeq \Delta$. We notice that in the generic case 
with finite inter-nodal scattering, namely with non zero $v^2$ and $w^2$, 
the saddle point equation remains the same apart form $u^2 \rightarrow u^2+2w^2+v^2$. 

The quasiparticle density of states $N(\epsilon)$, after the analytic continuation of $i\omega$ 
to the positive real axis, $i\omega \to \epsilon>0$,  turns out to be, for small $\epsilon$,  
\be
N(\epsilon) = \frac{2}{\pi^2 v_F v_\Delta} 
\left[ Q_0 \,\ln \frac{\displaystyle \Lambda}{\displaystyle \sqrt{\epsilon^2+Q_0^2}}
+ \frac{\epsilon}{2} \left(\frac{\pi}{2} - \tan^{-1}\frac{Q_0^2-\epsilon^2}{2\epsilon Q_0}\right)\right]. 
\label{DOS-Born}
\ee  
Therefore the density of states 
at the saddle-point acquires a finite value $N_0\sim Q_0/u^2$ at the chemical potential 
$\epsilon=0$, while, for $\epsilon\gg Q_0$, turns back to the linear dependence 
$N(\epsilon)\sim \epsilon$ as in the absence of disorder.

\subsection{Frequency dependent terms}

The first order term in the frequency is just 
\be
- \frac{\omega}{2} {\rm Tr}\left[Q\,s_3\,\mathcal{G}_0\right] = 
- \omega \frac{\pi}{2}\,\frac{N_0}{Q_0}\, \int d\mathbf{r}\, 
{\rm Tr}\left[Q(\mathbf{r})\,s_3\right].
\label{omega-1}
\ee
The second order term is readily found to be
\be
\frac{\omega^2}{8}{\rm Tr}\left[
\mathcal{G}_0\,\left\{Q,s_3\right\}\,\mathcal{G}_0\,\left\{Q,s_3\right\}\right]
= 
\frac{1}{32\pi^2} \frac{\omega^2}{v_F v_\Delta} \frac{1}{Q_0^2}\,
\int d\mathbf{r}\, 
{\rm Tr}\left[\{Q(\mathbf{r}),s_3\}^2\right].
\label{omega-2}
\ee
This term is negligible as compared to (\ref{omega-1}) for frequencies $\omega$ much smaller 
than $v_F\,v_\Delta\,N_0$. The latter is therefore the energy scale below which diffusion 
sets in at the mean field level, namely the so-called inverse relaxation time in the 
Born approximation, $1/\tau_0$. 
  
\subsection{Gradient expansion}

The Fourier transform of the current operator is 
\be
\J(\q) = \gamma_3\,\frac{1}{2}\left[\J_{1}(\q)+\J_{2}(\q)\right] 
+ \gamma_3\,\rho_3\,\frac{1}{2}\left[\J_{1}(\q)-\J_{2}(\q)\right],
\ee
where, at long wavelengths $\q\to 0$,  
\ba
\J_{1}(\q) &\rightarrow& \J_1 = \left(v_F\, ,\, i\,v_\Delta\,\tau_2\, s_1\right)\\
\J_{2}(\q) &\rightarrow& \J_2 = \left(i\,v_\Delta\,\tau_2\, s_1\, , \, v_F\right).
\ea
It is then easy to derive the following expression of the tensor product $\J\otimes\J^\dagger$
\bea
\J\otimes\J^\dagger &=& \left(
\begin{array}{cc}
J_\xi J_\xi^\dagger   & J_\xi J_\eta^\dagger \\
J_\eta J_\xi^\dagger  & J_\eta J_\eta^\dagger \\
\end{array}
\right) = 
\frac{1}{2}\left(v_F^2+v_\Delta^2\right) 
\left(
\begin{array}{cc}
1 & 0 \\
0 & 1 \\
\end{array}
\right)\nonumber \\
&& + \frac{1}{2}\rho_3 \,\left(v_F^2-v_\Delta^2\right) 
\left(
\begin{array}{cc}
1 & 0 \\
0 & -1 \\
\end{array}
\right) - i v_F v_\Delta \, \tau_2\, s_1\, \rho_3\, 
\left(
\begin{array}{cc}
0 & 1 \\
-1 & 0 \\
\end{array}
\right).
\label{JJ}
\eea
  
To evaluate the second order gradient correction in (\ref{NLSM-7}) we notice that  
\ba
&&\frac{1}{8}{\rm Tr}\left[{\cal{G}}_0\,\vec{\nabla}Q\cdot \J^\dagger\, 
{\cal{G}}_0\, \J\cdot \vec{\nabla}Q\right] 
=
\frac{1}{8}\sum_{i,j}\, {\rm Tr}\left[\partial_i Q\, \partial_j Q\, 
{\cal{G}}_0 \,J_j^\dagger\, {\cal{G}}_0\, 
J_i\right]\\
&&~~~= \frac{1}{16}\sum_{i,j}\, {\rm Tr}\left[\partial_i Q\, \partial_j Q\,\left( 
{\cal{G}}_0 \,J_j^\dagger\, {\cal{G}}_0\, 
J_i + {\cal{G}}_0 \,J_i^\dagger\, {\cal{G}}_0\, 
J_j\right)\right]\\
\ea
In the long wavelength limit
\ba
&&{\cal{G}}_0 \,J_j^\dagger\, {\cal{G}}_0\, 
J_i + {\cal{G}}_0 \,J_i^\dagger\, {\cal{G}}_0\, 
J_j = \frac{1}{V} \sum_\kap {\cal{G}}_{0\kap}\,J_j^\dagger\, 
{\cal{G}}_{0\kap}\, J_i   + 
{\cal{G}}_{0\kap}\,J_i^\dagger\, 
{\cal{G}}_{0\kap}\, J_j \\
&& = \delta_{ij}\, \frac{1}{4V}\sum_\kap\,\left[
{\cal{G}}_{1\kap}^2 +  {\cal{G}}_{2\kap}^2\right] \left[\left(v_F^2+v_\Delta^2\right) + 
\rho_3\, \left(v_F^2-v_\Delta^2\right)\, \left(\delta_{i\xi}-\delta_{i\eta}\right)
\right],  
\ea
since $
\sum_\kap\, {\cal{G}}_{1\kap}^2 - {\cal{G}}_{2\kap}^2 = 0$.
We find that   
\ba
&& \frac{1}{4V}\sum_\kap\,
{\cal{G}}_{1\kap}^2 +  {\cal{G}}_{2\kap}^2 = 
 \frac{1}{2V}\sum_\kap\,
{\cal{G}}_{1\kap}^2 \\
&& = 
\frac{1}{2}\int \frac{d\kap_\xi\, d\kap_\eta}{4\pi^2} \, 
\left(\frac{1}{v_F^2\kap_\xi^2 + v_\Delta^2 \kap_\eta^2 + Q_0^2}\right)^2 \\
&& = \frac{1}{8\pi v_F v_\Delta}\int_0^\infty\, dx^2 \, 
 \left(\frac{1}{x^2+ Q_0^2}\right)^2 = 
\frac{1}{8\pi v_F v_\Delta}\, \frac{1}{Q_0^2}.
\ea
The Drude spin-conductivity is defined by\cite{Lee}   
\be
\sigma = \frac{1}{4\pi^2}\frac{v_F^2+v_\Delta^2}{v_F v_\Delta},
\label{sigma}
\ee
so that the second order gradient correction can be written as   
\[
\frac{\pi}{16 \, Q_0^2}\,\sigma\, \int d\rr \, {\rm Tr}\left[\bnabla Q(\rr)\cdot 
\bnabla Q(\rr)\right] 
+ \frac{v_F^2-v_\Delta^2}{v_F^2+v_\Delta^2}
{\rm Tr}\left\{\rho_3\left[ 
\partial_\xi Q(\rr)\,\partial_\xi Q(\rr) \, -\,
\partial_\eta Q(\rr)\,\partial_\eta Q(\rr)\right]\right\}.
\]

\subsection{The final form of $S_{NL\sigma M}$}

Collecting all contributions which are second order in the gradient expansion 
and first order in the frequency we eventually obtain  
\bea
{\cal{S}}_{NL\sigma M} &=& 
\frac{\pi}{16\, Q_0^2}\,\sigma\, \int d\rr \, {\rm Tr}\left[\bnabla Q(\rr)\cdot 
\bnabla Q(\rr)\right]   \nonumber \\
&& +\frac{\pi}{16\, Q_0^2}\,\sigma\, \int d\rr \, \frac{v_F^2-v_\Delta^2}{v_F^2+v_\Delta^2}
{\rm Tr}\left\{\rho_3\left[ 
\partial_\xi Q(\rr)\,\partial_\xi Q(\rr) \, -\,
\partial_\eta Q(\rr)\,\partial_\eta Q(\rr)\right]\right\}\nonumber \\
&& - \omega\, \frac{\pi\, N_0}{2\,Q_0}  
\int d\rr \, {\rm Tr}\left[Q(\rr)\, s_3\right]\nonumber \\
&=& \sum_{i=1,2}\,  \int d\rr \, \frac{\pi}{16\, Q_0^2}\,\sigma\,
{\rm Tr}\left[\partial_\mu Q_i(\rr)\,\alpha^{(i)}_{\mu\nu} 
\partial_\nu Q_i(\rr)\right] - \omega\, \frac{\pi\, N_0}{2\,Q_0} \, 
{\rm Tr}\left[Q_i(\rr)\, s_3\right].
\label{NLSM-conventional-term}
\eea
Here we have defined a metric tensor for the pair 1 which includes nodes 
``1'' and ``$\overline{1}$'',
\[
\alpha^{(1)} = \frac{2}{v_F^2+v_\Delta^2}\, 
\left(
\begin{array}{cc}
v_F^2 & 0 \\
0 & v_\Delta^2 \\
\end{array}
\right),
\]
and for the pair 2, {\sl i.e.} for nodes ``2'' and ``$\overline{2}$'', 
\[
\alpha^{(2)} = \frac{2}{v_F^2+v_\Delta^2}\, 
\left(
\begin{array}{cc}
v_\Delta^2 & 0 \\
0 & v_F^2 \\
\end{array}
\right).
\]

In reality the above action is not the most general one allowed by symmetry. As we discussed 
earlier, the theory possesses two chiral abelian sectors, 
which actually occur in the singlet channels 
$\gamma_3\, \tau_2 \, s_1\, \rho_0$ and 
$\gamma_3\, \tau_2 \, s_1\, \rho_3$, see (\ref{chiral-30S}) 
and (\ref{chiral-33S}). In analogy with Ref.~[\onlinecite{NPB}], we expect that 
upon integrating out longitudinal fluctuations the following term would appear:
\be
\delta {\cal{S}}_{NL\sigma M} = \frac{\pi}{16^2\, Q_0^4}\, \sigma\, \Pi\, 
\sum_{i=1,2}\, \int d\rr \,{\rm Tr}\left[\gamma_3\,\tau_2\, s_1\, 
Q_i(\rr)\partial_\mu Q_i(\rr)\right]\,\alpha^{(i)}_{\mu\nu}\, 
{\rm Tr}\left[\gamma_3\,\tau_2\, s_1\, 
Q_i(\rr)\partial_\nu Q_i(\rr)\right],
\label{GADE}
\ee
with $\Pi\propto u^2$.

In conclusion the most general non-linear $\sigma$ model is given by 
\bea
{\cal{S}}_{NL\sigma M} &=& \sum_{i=1,2}\, \int d\rr \, 
\frac{\pi}{16\, Q_0^2}\,\sigma \,   
{\rm Tr}\left[\partial_\mu Q_i(\rr)\, \alpha^{(i)}_{\mu\nu}\,  
\partial_\nu Q_i(\rr)\right] - \omega\, \frac{\pi\, N_0}{2\,Q_0} \, 
{\rm Tr}\left[Q_i(\rr)\, s_3\right]\nonumber \\
&& + \frac{\pi}{16^2\, Q_0^4}\, \sigma\, \Pi\, 
{\rm Tr}\left[\gamma_3\,\tau_2\, s_1\, 
Q_i(\rr)\partial_\mu Q_i(\rr)\right]\,\alpha^{(i)}_{\mu\nu}\, 
{\rm Tr}\left[\gamma_3\,\tau_2\, s_1\, 
Q_i(\rr)\partial_\nu Q_i(\rr)\right].\label{conventional-NLSM}
\eea

\section{Failure of the conventional $\sigma$-model}

The non-linear $\sigma$-model (\ref{conventional-NLSM}) belongs to one 
of the known chiral $\sigma$-models encountered when two-sublattice symmetry 
holds, see Refs.~[\onlinecite{Gade,NPB,Luca}]. If we simply borrow known results, 
see e.g. Table I in Ref.~[\onlinecite{Luca}], we should expect that the $\beta$-function 
of the conductivity vanishes in the zero replica limit, $N\to 0$. That would imply 
absence of localization and persistence of diffusive modes. Moreover we should predict 
a quasiparticle density of states $N(\epsilon)$ which diverges like\cite{Damle,Mudry,luca2}
\be
\label{N(epsilon)}
N(\epsilon) \simeq \frac{1}{\epsilon}\, \exp\left[-A\sqrt[3]{ln\frac{B}{\epsilon} }\right],
\ee 
with $A$ and $B$ some positive constants. This is clearly suspicious since 
in the absence of disorder the density of states actually vanishes, 
$N(\epsilon)\sim \epsilon$. 

One is tempted to correlate the above suspicious result
with what is found for the elementary loops of 
the Wilson-Polyakov renormalization group approach. 
Here one integrates out iteratively degrees of freedom in a momentum shell from the 
highest cut-off $\Lambda$ to $\Lambda/s$, with $s>0$ eventually 
sent to infinity. In our case these 
fundamental loops for either nodes are given by   
\[
\frac{1}{\pi \sigma}\, \int_{\Lambda/s<|\kap|<\Lambda} \frac{d\kap}{4\pi^2}\, 
\frac{1}{\kappa_\mu\, \alpha^{(1,2)}_{\mu\nu}\, 
\kappa_\nu} = \ln s \equiv g\,\ln s,  
\]
and provide the definition of the dimensionless coupling constant $g$, which 
should necessarily be small to justify a loop-expansion in $g\ln s \ll 1$. 
In our case it turns out that $g=1$, making any loop expansion meaningless. 

This results is at odds with the standard non linear $\sigma$-models for 
disordered systems where $g\sim 1/\sigma \ll 1$ for weak disorder. 
Here, whatever weak is the disorder, yet $g=1$. This peculiar fact was 
originally discussed in Ref.~[\onlinecite{Nersesyan}], where the authors  
identified correctly the complete failure of the non-crossing approximation as a starting 
point of perturbation theory due to the absence of small control parameters. 
This might lead to the conclusion that the non-linear $\sigma$-model we have so far 
derived is useless in this problem, since it heavily relies upon the assumption that 
quantum fluctuations around the saddle-point solution can be controlled perturbatively. 

In the following Section we are going to show that the above   
conclusion is not correct and that one only needs to be more careful in 
deriving the proper non-linear $\sigma$-model action.

\section{Wess-Zumino-Novikov-Witten term}
Indeed we have not yet accomplished all our plan, as we still need to 
evaluate the imaginary part of the action (\ref{S-imaginary}). At leading order 
we can drop the frequency dependence of (\ref{S-imaginary}), hence  
\be
{\cal{S}}_{\Gamma} = - \frac{1}{4}{\rm Tr}\,\ln\left[iQ - {\cal{H}}_0\right] 
+ \frac{1}{4}{\rm Tr}\,\ln\left[-iQ - {\cal{H}}_0^\dagger\right].
\label{WZW-S-imaginary}
\ee 
It is more convenient to evaluate the variation of ${\cal{S}}_{\Gamma}$ along 
a massless path defined through 
\be
\delta Q(\rr)\, Q(\rr) + Q(\rr)\, \delta Q(\rr) = 0.
\ee
We find that 
\be 
\delta{\cal{S}}_{\Gamma} = - \frac{i}{4}{\rm Tr}\left[{\rm G}\, \delta Q\right]
+ \frac{i}{4}{\rm Tr}\left[{\rm G}^\dagger\, \delta Q\right],
\label{deltaS}
\ee
where
\be
{\rm G} = \left(iQ - {\cal{H}}_0\right)^{-1}.
\ee
We notice that, in the long wavelength limit,  
\be
\left(-iQ - {\cal{H}}_0^\dagger\right)\,\left(iQ - {\cal{H}}_0\right) 
\simeq {\cal{G}}_0^{-1} - \bnabla Q\cdot \J = 
{\cal{G}}_0^{-1} - \J^\dagger \cdot \bnabla Q,
\ee
and
\be
\left(iQ - {\cal{H}}_0\right)\, \left(-iQ - {\cal{H}}_0^\dagger\right)
\simeq {\cal{G}}_0^{-1} + \bnabla Q\cdot \J = 
{\cal{G}}_0^{-1} + \J^\dagger \cdot \bnabla Q,
\ee
hence
\ba
{\rm G} &=& \left[1-{\cal{G}}_0\,\bnabla Q\cdot \J\right]^{-1}\, {\cal{G}}_0\,
\left(-iQ - {\cal{H}}_0^\dagger\right)\\
&\simeq& \left[1 + {\cal{G}}_0\,\bnabla Q\cdot \J
+ {\cal{G}}_0\,\bnabla Q\cdot \J{\cal{G}}_0\,\bnabla Q\cdot \J\right]
\, {\cal{G}}_0\,
\left(-iQ - {\cal{H}}_0^\dagger\right),\\ 
{\rm G}^\dagger &=& 
\left[1+{\cal{G}}_0\,\bnabla Q\cdot \J\right]^{-1}\, {\cal{G}}_0\,
\left(iQ - {\cal{H}}_0\right)\\
&\simeq& 
\left[1 - {\cal{G}}_0\,\bnabla Q\cdot \J
+ {\cal{G}}_0\,\bnabla Q\cdot \J{\cal{G}}_0\,\bnabla Q\cdot \J\right]
{\cal{G}}_0\,
\left(iQ - {\cal{H}}_0\right).
\ea
The leading non-vanishing contribution to (\ref{deltaS}) derives from 
the second order gradient correction to $\rm G-G^\dagger$, which reads 
\be
{\rm G}-{\rm G}^\dagger \simeq -2\,i\, 
{\cal{G}}_0\,\bnabla Q\cdot \J{\cal{G}}_0\,\bnabla Q\cdot \J
\, {\cal{G}}_0\, Q.
\ee
Therefore  
\bea
\delta S_\Gamma &=& -\frac{1}{2}{\rm Tr}\left[
\delta Q\,{\cal{G}}_0\,\bnabla Q\cdot \J\; {\cal{G}}_0\,\bnabla Q\cdot \J 
\, {\cal{G}}_0\, Q\right]\nonumber \\
&=& -\frac{1}{2}{\rm Tr}\left[
\delta Q\,{\cal{G}}_0\,\bnabla Q\cdot \J\; {\cal{G}}_0\, \J^\dagger\cdot \bnabla Q\,  
 {\cal{G}}_0\, Q\right]
\eea    
The only term which contributes comes from the 
anti-symmetric component of the tensor $\J\otimes\J^\dagger$:
\be
\frac{1}{2}\left(J_i\,J_j^\dagger - J_j\,J_i^\dagger\right) = 
- i\,\epsilon_{ij}\, v_F\,v_\Delta\,\tau_2\,s_1\,\rho_3,
\ee
being $\epsilon_{ij}$ the Levi-Civita tensor, thus leading to 
\be
\delta S_\Gamma = i\,\Gamma\, \sum_{i,j}\,\epsilon_{i,j} {\rm Tr}\left[
\partial_j Q\, Q\, \delta Q\, \partial_i Q\, \tau_2\, s_1\, \rho_3\right]
\ee    
where 
\bea
\Gamma &=& \frac{1}{4V}\sum_\kap v_F\,v_\Delta \left[ {\cal{G}}_{1\kap}^3 
+ {\cal{G}}_{2\kap}^3\right] \nonumber \\
&=& \frac{1}{8\pi}\int_0^\infty \, dx^2 \left(\frac{1}{x^2+Q_0^2}\right)^3 
= \frac{1}{16\pi}\frac{1}{Q_0^4}.
\label{anomaly}
\eea 
Introducing a fictitious coordinate which parametrizes the massless path, we 
finally get
\bea
{\cal{S}}_\Gamma &=& i\,\frac{1}{12\pi}\,\frac{1}{4Q_0^6}\,
\int d^3 \rr \; \epsilon_{\mu\nu\eta}\, 
{\rm Tr}\left[\tau_2\,s_1\,\rho_3\, 
Q(\rr)\,\partial_\mu Q(\rr) \; Q(\rr)\,\partial_\nu Q(\rr)\;      
Q(\rr)\,\partial_\eta Q(\rr)  \right]\label{final-WZW}\\
&=& i\,\frac{1}{12\pi}\,\frac{1}{4Q_0^6}\,
\int d^3 \rr \; \epsilon_{\mu\nu\eta}\, 
{\rm Tr}\left[\tau_2\,s_1\, 
Q_1(\rr)\,\partial_\mu Q_1(\rr) \; Q_1(\rr)\,\partial_\nu Q_1(\rr)\;      
Q_1(\rr)\,\partial_\eta Q_1(\rr)  \right]\nonumber \\
&& - i\,\frac{1}{12\pi}\,\frac{1}{4Q_0^6}\,
\int d^3 \rr \; \epsilon_{\mu\nu\eta}\, 
{\rm Tr}\left[\tau_2\,s_1\, 
Q_2(\rr)\,\partial_\mu Q_2(\rr) \; Q_2(\rr)\,\partial_\nu Q_2(\rr)\;      
Q_2(\rr)\,\partial_\eta Q_2(\rr)  \right].\nonumber 
\eea 
This actually represents opposite WZW terms for each of the 
two pairs of nodes, namely for each of the two 
independent U(1)$\times$SU($4N$) $\sigma$-models. It is clear that if the 
disorder coupled the two pairs of nodes, we would be forced to identify  
$Q_1(\rr)$ with $Q_2(\rr)$, so that the WZW term 
would cancel in that case. To make (\ref{final-WZW}) more explicit we remind 
that, for $a=1,2$, 
\be
Q_a(\rr) = T_a(\rr)^{-2}\, s_3\, Q_0 = s_3\, Q_0\, T_a(\rr)^2,
\label{Z1}
\ee
where 
\be
T^2_a(\rr) = \frac{1}{2}\left(\gamma_0+\gamma_3\right)\, V^2_a(\rr)
+ \frac{1}{2}\left(\gamma_0-\gamma_3\right)\, c^t\, \left[V^{-2}_a(\rr)\right]^t\, c,
\label{Z2}
\ee
and 
\be
V_a^{2}(\rr) = U\, \left[ \frac{1}{2}\left(s_0+s_3\right)\,
{\rm e}^{i\sqrt{\frac{\pi}{N}}\,\phi_a(\rr)}\, g_a(\rr) + 
\frac{1}{2}\left(s_0-s_3\right)\,
{\rm e}^{-i\sqrt{\frac{\pi}{N}}\,\phi_a(\rr)}\, g_a(\rr)^\dagger\right]\, U^\dagger,
\label{Z3}
\ee
$U$ being defined in Eq.~(\ref{U-unitary}). By means of (\ref{Z1}), 
(\ref{Z2}) and (\ref{Z3}), the expression (\ref{final-WZW}) can be finally written as 
\bea
{\cal{S}}_\Gamma &=& i\,\frac{1}{12\pi}\,
\int d^3 \rr \; \epsilon_{\mu\nu\eta}\, 
{\rm Tr}\left[ 
g_1(\rr)^\dagger\,\partial_\mu g_1(\rr) \; g_1(\rr)^\dagger\,\partial_\nu g_1(\rr)\;      
g_1(\rr)^\dagger\,\partial_\eta g_1(\rr)  \right]\nonumber \\
&& - i\,\frac{1}{12\pi}\,
\int d^3 \rr \; \epsilon_{\mu\nu\eta}\, 
{\rm Tr}\left[ 
g_2(\rr)^\dagger\,\partial_\mu g_2(\rr) \; g_2(\rr)^\dagger\,\partial_\nu g_2(\rr)\;      
g_2(\rr)^\dagger\,\partial_\eta g_2(\rr) \right],
\label{WZW-end}
\eea
appropriate for two WZW models SU$(4N)_1$. 

Let us express all other terms in the action by means of $\phi_a$ and $g_a$. 
First of all the density of state operator becomes
\be
\frac{1}{Q_0}\,{\rm Tr}\left[s_3\, Q(\rr)\right] =
{\rm Tr}\left[T^2(\rr)\right] = 2\sum_{a=1}^2\, 
\,{\rm e}^{i\,\sqrt{\frac{\pi}{N}}\,\phi_a(\rr)}
{\rm Tr}\left[g_a(\rr)\right] +
{\rm e}^{-i\,\sqrt{\frac{\pi}{N}}\,\phi_a(\rr)}
{\rm Tr}\left[g_a(\rr)^\dagger\right].  
\label{DOS-g}
\ee
Moreover one readily finds that 
\[
\frac{1}{Q_0^2}\,{\rm Tr}\left[\partial_\mu Q_a(\rr)\, \alpha^{(a)}_{\mu\nu}\,
\partial_\nu Q_a(\rr) \right] = 
16\pi \,\partial_\mu \phi_a(\rr) \, \alpha^{(a)}_{\mu\nu}\,\partial_\mu \phi_a(\rr)
+ 4 {\rm Tr}\left[\partial_\mu g_a(\rr)\, \alpha^{(a)}_{\mu\nu}\,
\partial_\nu g_a(\rr)^\dagger \right],
\]
and 
\[
\frac{1}{Q_0^4}{\rm Tr}\left[\gamma_3\,\tau_2\, s_1\, 
Q_a(\rr)\partial_\mu Q_a(\rr)\right]\,\alpha^{(a)}_{\mu\nu}\, 
{\rm Tr}\left[\gamma_3\,\tau_2\, s_1\, 
Q_a(\rr)\partial_\nu Q_a(\rr)\right] = 
\frac{\pi\left(16N\right)^2}{N} \, 
\partial_\mu \phi_a(\rr) \, \alpha_{\mu\nu}\,\partial_\mu \phi_a(\rr).
\]

In conclusion the action expressed in terms of the fundamental fields and including 
the WZW term reads  

\bea
{\cal{S}}_{WZW} &=& \sum_{a=1,2}\, \int d\rr \, 
\frac{\pi}{4}\,\sigma \,   
{\rm Tr}\left[\partial_\mu g_a(\rr)\, \alpha^{(a)}_{\mu\nu}\,  
\partial_\nu g_a(\rr)^\dagger \right] 
+ \pi^2\,\sigma\, \left(1+N\Pi\right) 
\,\partial_\mu \phi_a(\rr) \, \alpha^{(a)}_{\mu\nu}\,\partial_\mu \phi_a(\rr)\nonumber \\
&& - 
\omega\, \pi\, N_0 \, \left\{
{\rm e}^{i\,\sqrt{\frac{\pi}{N}}\,\phi_a(\rr)}
{\rm Tr}\left[g_a(\rr)\right] +
{\rm e}^{-i\,\sqrt{\frac{\pi}{N}}\,\phi_a(\rr)}
{\rm Tr}\left[g_a(\rr)^\dagger\right]\right\}  
+ S^{(a)}_\Gamma,
\label{WZW-NLSM}
\eea
with
\be
S^{(a)}_\Gamma = \pm 
i\,\frac{1}{12\pi}\,
\int d^3 \rr \; \epsilon_{\mu\nu\eta}\, 
{\rm Tr}\left[ 
g_a(\rr)^\dagger\,\partial_\mu g_a(\rr) \; g_a(\rr)^\dagger\,\partial_\nu g_a(\rr)\;      
g_a(\rr)^\dagger\,\partial_\eta g_a(\rr)  \right], 
\ee
where the plus refers to $a=1$ and the minus to $a=2$.

\section{Consequences of the WZW term}
\label{consequences}

We have just shown that the actual field theory which describes $d$-wave 
superconductors in the presence of a disorder which only permits intra-node scattering 
processes is not a conventional non linear $\sigma$-model but instead it represents   
two decoupled U(1)$\times$SU$(4N)_1$ WZW models. 
Moreover if, for instance, we consider pair 1, then 
\ba
&&\frac{\pi}{4}\,\sigma\, \frac{2}{v_F^2+v_\Delta^2} \, 
\int d^2\rr \; v_F^2 {\rm Tr}\left[\partial_\xi g_a(\rr)\,   
\partial_\xi g_a(\rr)^\dagger \right] 
+ v_\Delta^2 {\rm Tr}\left[\partial_\eta g_a(\rr)\,   
\partial_\eta g_a(\rr)^\dagger \right] \\
&& = \frac{\pi}{4}\,\frac{1}{4\pi^2}\, \frac{v_F^2+v_\Delta^2}{v_Fv_\Delta}\,
\frac{2}{v_F^2+v_\Delta^2} \, 
\int d^2\rr \; v_F^2 {\rm Tr}\left[\partial_\xi g_a(\rr)\,   
\partial_\xi g_a(\rr)^\dagger \right] 
+ v_\Delta^2 {\rm Tr}\left[\partial_\eta g_a(\rr)\,   
\partial_\eta g_a(\rr)^\dagger \right] \\
&& = \frac{1}{8\pi}\, 
\int \frac{d^2\rr}{v_Fv_\Delta} \; v_F^2 {\rm Tr}\left[\partial_\xi g_a(\rr)\,   
\partial_\xi g_a(\rr)^\dagger \right] 
+ v_\Delta^2 {\rm Tr}\left[\partial_\eta g_a(\rr)\,   
\partial_\eta g_a(\rr)^\dagger \right],
\ea
which, upon the change of variable $\xi\to v_F\xi$ and $\eta\to v_\Delta \eta$, 
shows that each WZW model is right at its fixed point. Hence there is no 
ambiguity in the zero replica limit.

Now we can draw some consequences of what we have found. 
The first is that the average value of the density of states (DOS) at the chemical potential 
stays zero, as in the absence of disorder and contrary to the Born approximation. In particular 
the dimension of the density of states operator in the zero replica limit $N\to 0$ is 
\[
\Delta_Q = \frac{N-1}{N} + \frac{1}{N(1+N\Pi)} \to 1 - \Pi,
\]
while the dimension of the frequency is $[\omega] =2-(1 - \Pi)=1 + \Pi$.
This implies that the DOS at finite frequency behaves as 
\be
N(\omega) \sim |\omega|^{\frac{1-\Pi}{1+\Pi}},
\label{N(omega)}
\ee
in agreement with Ref.~[\onlinecite{Nersesyan}]. 

Notice the (\ref{N(omega)}) stems from the fact that the WZW terms modifies
the non-linear $\sigma$-model results leading to Eq. (\ref{N(epsilon)}) in two
ways: i)it makes $\Pi$ unrenormalized as well as $\sigma$, ii) it adds the further
contribution one to the dimension of the density of states operator.
The only difference between the 1d
mapping and the non-linear $\sigma$-model results for the density of states are the constant
factors fixed by the range of validity.
1d mapping: $N(\omega) \sim 1/(v_Fv_\Delta) \omega (\omega/\Delta)^{-2\Pi/(1+\Pi)}$ with
$\omega<\Delta$; non-linear $\sigma$-model: $N(\omega) \sim N_0 (\omega\tau_0)^{(1-\Pi)/(1+\Pi)}$
with $\omega<1/\tau_0 $, $N_0$ being the saddle point value of the density of states. At leading ording in $u^2$ the
matching of the two expressions is provided by $N_0\tau_0 \sim 1$.

The second consequence concerns the transport properties. We have 
shown that within the Born approximation a finite spin-conductivity arises. 
Is this result still true beyond that approximation? 
The renormalization group analysis which identifies the spin conductivity
with the  coupling of the non-linear $\sigma$-model would say that $\sigma_s$ 
stays unrenormalized to its fixed point value. Can we understand this in the 
1d mapping language?. 

In order to answer this question, 
we introduce a uniform spin vector potential within the action, 
which in our path-integral approach has the form
\be
\AAA = \AAA_0 \, \sigma_3 + \AAA_1 \, \sigma_3\, s_1.
\label{spin-vector-potential}
\ee
The action becomes now a functional of $\AAA$, {\sl i.e.} ${\cal{S}}\rightarrow 
{\cal{S}}(\AAA)$, and the spin-conductivity turns out to be given by
\be
\sigma_s = \frac{1}{2\pi}\left(
\frac{\partial^2 \ln {\cal{Z}}(\AAA)}{\partial A_{0,\xi}^2}
-\frac{\partial^2 \ln {\cal{Z}}(\AAA)}{\partial A_{1,\xi}^2}\right)_{\AAA=0}
= \frac{1}{2\pi} \left(
\frac{\partial^2 \ln {\cal{Z}}(\AAA)}{\partial A_{0,\eta}^2}
-\frac{\partial^2 \ln {\cal{Z}}(\AAA)}{\partial A_{1,\eta}^2}\right)_{\AAA=0},
\ee
where 
\[
{\cal{Z}}(\AAA) = \int \, DQ(\rr)\; {\rm}^{-{\cal{S}}(\AAA)} 
\]
is the generating functional in the presence of $\AAA$.
It is possible to show that at second order in $\AAA$ one needs just to make the 
following substitution in the action
\be
\partial_\mu Q(\rr) \rightarrow \partial_\mu Q(\rr) + \frac{i}{2}\left[Q(\rr),A_\mu\right].
\label{intermediate}
\ee
Upon the action of (\ref{U-unitary}) 
\[
\AAA \rightarrow U^\dagger\, \AAA\, U = \AAA_0 \, \sigma_3 + 
\AAA_1 \, \sigma_3\, \tau_2\, s_3,
\]
and we can show that (\ref{intermediate}) implies the following transformation law of the matrix field $G(\rr)$ in the presence of $\AAA$
\be
G(\rr) \rightarrow 
{\rm e}^{-\frac{i}{2}\int d\rr' \cdot \,
\left(\AAA_0(\rr')\,\sigma_3 - \AAA_1(\rr')\, \tau_2\,\sigma_3\, \right)}\; 
G(\rr)\;{\rm e}^{\frac{i}{2}\int d\rr' \cdot\, 
\left(\AAA_0(\rr')\,\sigma_3\, + \AAA_1(\rr')\, \tau_2\, \sigma_3\, \right)}.
\label{g-transformation}
\ee  
In order to better understand the role of $\mathbf{A}$ it is 
convenient to translate the WZW action into 
the language of the underneath free one-dimensional fermions.  
One may identify 
the component $G_{1;\alpha\beta}(\rr)$ of the matrix field for pair ``1'', 
where $\alpha$ and $\beta$ run over $4N$ indices, with 
\be
G_{1;\alpha\beta}(\rr) \rightarrow i\, 
\Psi^\dagga_{1;R\, \alpha}(\rr) \,\Psi^\dagger_{1;L\, \beta}(\rr),
\label{g-RL}
\ee
being $\Psi_{1;\alpha\,R}(\rr)$ and $\Psi_{1;\alpha\,L}(\rr)$ right and left moving 
one dimensional Fermi fields, respectively, and the two component vector $\rr$ playing 
the role of space and time coordinates. Since pair ``2'' has the opposite WZW term 
of pair ``1'', it is more appropriate to define 
\be
G_{2;\alpha\beta}(\rr) \rightarrow -i\, 
\Psi^\dagga_{2;L\, \alpha}(\rr) \,\Psi^\dagger_{2;R\, \beta}(\rr),
\label{g2-RL}
\ee
which formally yields to equal WZW terms.  
Then (\ref{g-transformation}) implies for 
the fermions the transformation laws
\ba 
&&\Psi^\dagga_{1;R}(\rr) \rightarrow 
{\rm e}^{-\frac{i}{2}\int d\rr' \cdot \,
\left(\AAA_0(\rr')\,\sigma_3 - \AAA_1(\rr')\, \tau_2\,\sigma_3\, \right)}\;
\Psi^\dagga_{1;R}(\rr),\\
&&\Psi^\dagga_{1;L}(\rr) \rightarrow 
{\rm e}^{-\frac{i}{2}\int d\rr' \cdot \,
\left(\AAA_0(\rr')\,\sigma_3 + \AAA_1(\rr')\, \tau_2\,\sigma_3\, \right)}\;
\Psi^\dagga_{1;L}(\rr),\\
&&\Psi^\dagga_{2;R}(\rr) \rightarrow 
{\rm e}^{-\frac{i}{2}\int d\rr' \cdot \,
\left(\AAA_0(\rr')\,\sigma_3 + \AAA_1(\rr')\, \tau_2\,\sigma_3\, \right)}\;
\Psi^\dagga_{2;R}(\rr),\\
&&\Psi^\dagga_{2;L}(\rr) \rightarrow 
{\rm e}^{-\frac{i}{2}\int d\rr' \cdot \,
\left(\AAA_0(\rr')\,\sigma_3 - \AAA_1(\rr')\, \tau_2\,\sigma_3\, \right)}\;
\Psi^\dagga_{2;L}(\rr),
\ea
which we may interpret as if $\AAA_0$ couples to the spin-density operator 
\be
\sum_{i=1,2} \Psi^\dagger_{i;R}(\rr) \,\sigma_3\,\Psi^\dagga_{i;R}(\rr) 
+ \Psi^\dagger_{i;L}(\rr) \,\sigma_3\,\Psi^\dagga_{i;L}(\rr),
\label{A0-fermion}
\ee
while $\AAA_1$ to the spin-current operator 
\be
\Psi^\dagger_{1;R}(\rr) \,\sigma_3\,\tau_2\,\Psi^\dagga_{1;R}(\rr) 
- \Psi^\dagger_{1;L}(\rr) \,\sigma_3\,\tau_2\,\Psi^\dagga_{1;L}(\rr)
- \Psi^\dagger_{2;R}(\rr) \,\sigma_3\,\tau_2\,\Psi^\dagga_{2;R}(\rr) 
+ \Psi^\dagger_{2;L}(\rr) \,\sigma_3\,\tau_2\,\Psi^\dagga_{2;L}(\rr).
\label{A1-fermion}
\ee 
Since the fermions are free, 
apart from the abelian sector which is not coupled to $\AAA$, the susceptibility towards 
$\AAA_0$ is the same as that towards $\AAA_1$, hence the spin-conductivity would seem to 
vanish at $\omega$ strictly equal to zero, again unlike what we found within the 
Born approximation. 
Actually one has to be more careful in drawing such a conclusion. Let us suppose to 
do the same calculation at finite frequency $\omega$ and afterwards send $\omega\to 0$. 
A frequency $\omega\not = 0$ plays the role of an explicit dimerization in the 
one dimensional fermionic problem:
\be
-\pi\,\omega\, N_0\,\sum_{i=1,2} {\rm Tr}\left[G^\dagga_i(\rr)+
G_i^\dagger(\rr)\right]
= -i\,\pi\,\omega\, N_0\, \sum_{i,\alpha}\, 
\Psi^\dagger_{i;R\,\alpha}(\rr)\Psi^\dagga_{i;L\,\alpha}(\rr)
-\Psi^\dagger_{i;L\,\alpha}(\rr)\Psi^\dagga_{i;R\,\alpha}(\rr).
\label{dimerization}
\ee
It is straightforward to show that in the presence 
of a finite dimerization the current-current susceptibility is finite and practically 
equal to that one in the absence of dimerization, while the 
density-density susceptibility is zero. This discontinuous behavior at $\omega =0$ 
as opposed to $\omega\to 0$ is again a manifestation of the chiral anomaly which plays 
such an important role in this problem\cite{Nersesyan}. Since it is physically 
more appropriate to identify the spin conductance through the $\omega\to 0$ limit, 
we conclude that, in spite of the vanishing DOS, the spin conductivity is 
finite. In other words, in spite of the fact that the DOS is vanishing at the chemical 
potential, namely that quasiparticle motion is undamped hence remains ballistic, yet 
the spin conductivity acquires a finite value in agreement with the Drude approximation.

\section{Inter-node scattering processes}

So far we have just considered the role of impurity scattering within each node. Now 
let us extend our analysis by including also inter-node scattering processes. 
Upon integrating out the most general disorder, we find two additional terms. The 
first describes opposite node scattering processes, and it reads:
\be
\delta {\cal{S}}_{imp}^{I} = - v^2\, 
\int d\rr \; 
\left[\overline{\Psi}_1(\rr)\,\Psi_{\overline{1}}(\rr)\right]
\left[\overline{\Psi}_{\overline{1}}(\rr)\,\Psi_1(\rr)\right] + 
\left[\overline{\Psi}_2(\rr)\,\Psi_{\overline{2}}(\rr)\right]
\left[\overline{\Psi}_{\overline{2}}(\rr)\,\Psi_2(\rr)\right].
\label{v-v}
\ee

The second derives from the impurity scattering processes which couple the two pairs of 
nodes, and it is given by 
\bea
\delta {\cal{S}}_{imp}^{II} &=& 
- w^2\, 
\int d\rr \; 
\left[\overline{\Psi}_1(\rr)\,\Psi_2(\rr)\right]
\left[\overline{\Psi}_2(\rr)\,\Psi_1(\rr)\right]
+ \left[\overline{\Psi}_{\overline{1}}(\rr)\,\Psi_{\overline{2}}(\rr)\right]
\left[\overline{\Psi}_{\overline{2}}(\rr)\,\Psi_{\overline{1}}(\rr)\right]
\nonumber \\
&&~~~~\,+\, 
\left[\overline{\Psi}_1(\rr)\,\Psi_{\overline{2}}(\rr)\right]
\left[\overline{\Psi}_{\overline{2}}(\rr)\,\Psi_1(\rr)\right]
+
\left[\overline{\Psi}_{\overline{1}}(\rr)\,\Psi_{2}(\rr)\right]
\left[\overline{\Psi}_{2}(\rr)\,\Psi_{\overline{1}}(\rr)\right]. 
\label{w-w}
\eea 
As before we can decouple the four-fermion terms by introducing auxiliary 
Hubbard-Stratonovich fields. Since these fields are 
expected to be massive, we can further expand the action up to second order in those fields 
after integrating out fermions. We finally integrate on the auxiliary massive fields.
The net result is still an action for the $Q$ matrices only 
which includes now the additional terms:
\be
\delta {\cal{S}}_{imp}^{I} = -\, \lambda^{I}\, \frac{N_0}{Q_0} \,  
\int d\rr\; {\rm Tr}\left[Q(\rr)\,\gamma_1\, Q(\rr)\,\gamma_1\right],
\label{SI}
\ee
and   
\be
\delta {\cal{S}}_{imp}^{II} = -\, \lambda^{II}\, \frac{N_0}{Q_0} \,  
\int d\rr\; \sum_{i=0,1}\, 
{\rm Tr}\left[Q(\rr)\,\rho_1\, \gamma_i \, Q(\rr)
\,\rho_1\, \gamma_i\right],
\label{SII}
\ee
where 
\[
\lambda^{I} \sim \frac{v^2}{u^2},\;\;
\lambda^{II} \sim \frac{w^2}{u^2}.
\]
The first term (\ref{SI}) tends to lock $G_1=G_{\overline{1}}$ and 
$G_2=G_{\overline{2}}$, while (\ref{SII}) locks $G_1=G_2$. 

When opposite node scattering is added, still leaving pairs of opposite
nodes uncoupled, only the symmetric Q-combinations of opposite nodes stay massless, 
the $\Pi$-term disappears and the coset for soft modes is Sp(2N) for each pair of
nodes. The $\beta$-function is vanishing only because of the contribution of
the WZW term, and density of states vanishes with a universal exponent in agreement
with the known results\cite{Nersesyan,Fendley,Fukui}.
In the absence of isotropic scattering the vanishing of the $\beta$ function still indicates
that the spin and heat conductivities have a metallic behavior. 

Finally, in the generic case in which all nodes are coupled only the four nodes 
symmetric combination of the Q's is required for the soft modes. The coset is 
again Sp(2N), but now it represents degrees of freedom coming from all nodes.
The two WZW terms are written in terms of the same Q and they cancels since they have opposite
sign. The action then reduce to the $S_{NL\sigma M}$ as derived by
[\onlinecite{Fisher}].

\subsection{Scaling analysis of the general model}
\begin{figure}
\centerline{{\includegraphics[width=13cm]{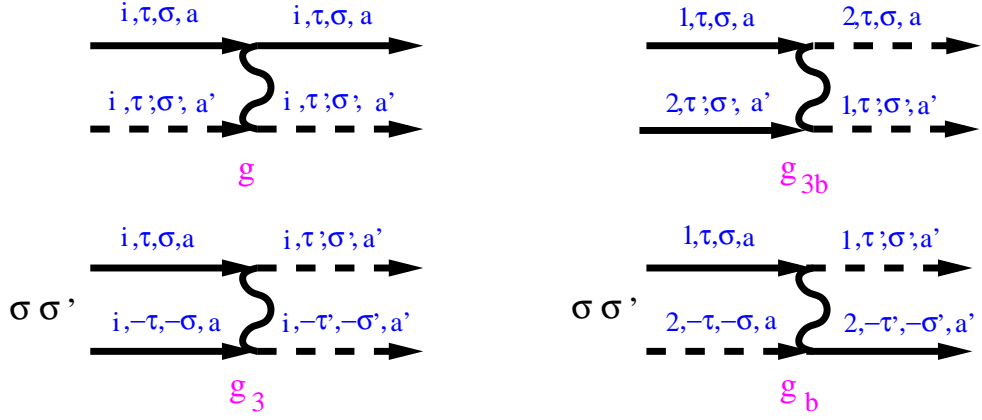}}}
\caption{Interaction vertices for the most general disorder. 
Solid(dashed) lines refer to right(left) moving fermions. The label 
$i=1,2$ refers to the two pairs of opposite nodes, $\tau,\tau'=\pm 1$ to the 
two Nambu components, $\sigma,\sigma'=\uparrow,\downarrow$ to the spin and 
$a,a'=1,\dots,N$ to the replicas. The symbol $\sigma\,\sigma'$ in front of 
the $g_3$ and $g_{3b}$ coupling constants means +1 if 
$\sigma=\sigma'$ and -1 otherwise.}
\label{couplings}
\end{figure}
In order to elucidate the role of inter-node scattering processes, it is convenient to 
transform the 2 dimensional non-linear $\sigma$-model into a 1+1 dimensional model of 
interacting fermions. 

We represent the matrix fields for pair ``1'' according to:    
\be
\begin{array}{lll}
G^\dagga_{1;\tau\,\sigma\, a \, , \, \tau'\,\sigma'\, a'}&=& i\, 
\Psi^\dagga_{1;R,\tau\,\sigma\, a}\, \Psi^\dagger_{1;L,\tau'\,\sigma'\, a'},\\
G^\dagger_{1;\tau\,\sigma\, a \, , \, \tau'\,\sigma'\, a'}&=&-i\, 
\Psi^\dagga_{1;L,\tau\,\sigma\, a}\, \Psi^\dagger_{1;R,\tau'\,\sigma'\, a'},\\
G^\dagga_{\overline{1};\tau\,\sigma\, a \, , \, \tau'\,\sigma'\, a'}&=&
\left(\tau_1\,\sigma_2\, G^\dagger_1\,\sigma_2\,\tau_1\right)_{\tau'\,\sigma'\, a'
\, , \,\tau\,\sigma\, a} \\
~~~&=& -i\, \sigma\,\sigma'\, 
\Psi^\dagga_{1;L,-\tau'\,-\sigma'\, a'}\, \Psi^\dagger_{1;R,-\tau\,-\sigma\, a}\\
G^\dagger_{\overline{1};\tau\,\sigma\, a \, , \, \tau'\,\sigma'\, a'}&=&
i\, \sigma\,\sigma'\,
\Psi^\dagga_{1;R,-\tau'\,-\sigma'\, a'}\, \Psi^\dagger_{1;L,-\tau\,-\sigma\, a}.
\end{array}
\label{GvsPsi}
\ee
As we mentioned, since pair ``2'' has the opposite WZW term of 
pair ``1'', it is convenient to define 
$G^\dagga_{2;\tau\,\sigma\, a \, , \, \tau'\,\sigma'\, a'}=-i\, 
\Psi^\dagga_{2;L,\tau\,\sigma\, a}\, \Psi^\dagger_{2;R,\tau'\,\sigma'\, a'}$.
In such a way the two WZW terms become equal leaving no ambiguity 
in mapping the non-linear $\sigma$-model onto a one-dimensional model of 
interacting electrons with the interaction vertices drawn in Fig.~\ref{couplings}. 
We notice that the coupling $g$ in Fig.~\ref{couplings} derives from the two terms 
in Eq.~(\ref{GADE}).
The bare values of the coupling constants are approximately 
\ba
g^{(0)} &\simeq& u^2,\\
g_3^{(0)} &\simeq& v^2,\\
g_b^{(0)} &=& g_{3b}^{(0)}\simeq w^2. 
\ea
Generally $u^2\geq w^2 \geq v^2$, the equality holding only for short range impurity 
potential. 
An important observation is that two-loop corrections 
to the renormalization group (RG) equations vanish in the zero 
replica limit, hence the RG equations valid up to two-loops 
are found to be:
\ba
\frac{dg}{d\ln s} &=& g_3^2 + g_b^2+g_{3b}^2,\\  
\frac{dg_{3}}{d\ln s} &=& 4\,g\,g_3 + 4\,g_b\,g_{3b},\\
\frac{dg_{b}}{d\ln s} &=& 2\,g\,g_b + 2\,g_{3}\,g_{3b},\\
\frac{dg_{3b}}{d\ln s} &=& 2\,g_3\,g_b + 2\,g_{3b}\,g,
\ea
where $s\to \infty$ is the scaling parameter. As discussed in Ref.~[\onlinecite{Zirnbauer}], 
the velocity anisotropy does not 
enter the RG equations, which remains true at least up to two loops in our fermionic 
replica trick approach. 
It is convenient to define $g_{\pm}=g_{3b}\pm g_{b}$, so that 
\ba
\frac{dg}{d\ln s}     &=& g_3^2 + \frac{1}{2}\,\left(g_+^2+g_{-}^2\right),\\  
\frac{dg_{3}}{d\ln s} &=& 4\,g\,g_3 + \left(g_+^2-g_{-}^2\right),\\
\frac{dg_{+}}{d\ln s} &=& 2\,\left(g+g_3\right)\, g_+,\\
\frac{dg_{-}}{d\ln s} &=& 2\,\left(g-g_3\right)\, g_-.
\ea
Given the appropriate bare values of the amplitudes, one readily recognizes that 
the RG flow maintains the initial condition $g_-=0$, hence the scaling equations 
reduce to 
\bea
\label{dg}
\frac{dg}{d\ln s}     &=& g_3^2 + \frac{1}{2}\,g_{+}^2,\\  
\frac{dg_{3}}{d\ln s} &=& 4\,g\,g_3 + g_{+}^2,\\
\frac{dg_{+}}{d\ln s} &=& 2\,\left(g+g_3\right)\, g_+.
\label{dg+}
\eea
The RG equations with the appropriate initial conditions always flow to strong coupling with 
\[
g_3\, \sim\,  2\, g\,  \sim\,  g_+ \,  \to\,  \frac{1}{3}\;\frac{1}{\ln s_c - \ln s},
\]
where $s_c$ can be interpreted as the correlation length of the modes which acquire a 
mass gap $E_c$ by the interaction, with  
\be
E_c \simeq \frac{\displaystyle \sqrt{v_F v_\Delta}}{\displaystyle s_c}.
\label{Ec-sc}
\ee
Therefore the two pairs of nodes get strongly coupled, 
in agreement with Refs.~[\onlinecite{Nersesyan,Zirnbauer,Fendley}]. 

\subsection{Strong coupling analysis} 
In order to gain further insight into the strong coupling phase towards which RG flows, 
let us consider 
the case of a single replica $N=1$. For further simplification it is convenient to 
adopt the same approach as in Refs.~\onlinecite{Nersesyan,Fendley,Zirnbauer,Fisher} 
and neglect the role of the opposite frequencies, which amounts to drop the 
$\tau$-label. The model thus reduces to two interacting chains of spinful fermions, each chain 
representing a pair of nodes. 
The coupling $g$ of Fig.~\ref{couplings} only couples to the charge sector and makes 
the intra-chain umklapp, the coupling $g_3$, a relevant perturbation which opens a charge gap 
on each chain. Therefore the model is equivalent to two coupled 
spin-1/2 chains. If we denote 
by $\vec{n}_{1(2)}(x)$ and $\epsilon_{1(2)}(x)$ respectively the staggered magnetization 
and the dimerization of chain 1(2), the coupling among the chains is ferromagnetic and 
given by 
\[
w^2 \,\int \, dx\, \epsilon_1(x)\,\epsilon_2(x)\, - \, 
4\, \vec{n}_1(x)\cdot \vec{n}_2(x). 
\]
As shown in Ref.~\onlinecite{Alexei}, this model is equivalent to an SO(4) Gross-Neveau 
which turns out to be fully massive or, equivalently, by four two-dimensional  
off-critical classical Ising models, three ordered and one disordered,  
$\langle \sigma_1 \rangle = \langle \sigma_2 \rangle = \langle \sigma_3 \rangle
= \langle \mu_4 \rangle = \sigma \not = 0$, where $\sigma_i$ and $\mu_i$, $i=1,\dots,4$, 
are order and 
disorder parameters, respectively. The ground state is rigid to an external 
magnetic field and to a spin vector potential opposite for the two chains, hence the 
conductivity is zero. As we discussed, a finite frequency amounts to add a term
\[
\omega\, \int\, dx\, \epsilon_1(x) + \epsilon_2(x) 
\propto \omega\, \int\, dx\, \sigma_1(x)\, \sigma_2(x)\, \sigma_3(x)\, 
\sigma_4(x) \simeq \omega\, \sigma^3 \, \int\, dx\, \sigma_4(x), 
\]
which actually plays the role of an external magnetic field acting on the 
fourth disordered Ising copy. The net result is that 
$\langle \sigma_4(x) \rangle_{\omega\not = 0} \sim \omega$, which in turns mean that the 
DOS remains linear in frequency. Even in the presence of an explicit dimerization, the 
susceptibilities towards a magnetic field or towards a spin vector potential opposite 
for the two chains still vanish. Were these results valid for any $N$, even for $N\to 0$, 
we should conclude that {\sl i)} the model is indeed insulating; {\sl ii)} the DOS 
is linear in frequency, in full agreement with Ref.~\onlinecite{Fisher2}.

\subsection{Identification of relevant energy scales}

The previous analysis of the $N=1$ model shows that the vanishing of thermal and 
spin conductivities in the model for a disordered $d$-wave superconductor translates 
in the language of the effective one-dimensional fermionic model 
into the existence of a finite spin-gap. The correlation length associated with this 
gap should then represent the localization length of the Bogoliubov quasiparticles. 
We may estimate this correlation length as the scale $s_c$ at which 
the RG equations (\ref{dg}-\ref{dg+}) encounter a singularity. In addition we 
may introduce the mass gap $E_c$ through Eq.~(\ref{Ec-sc}) which can be identified 
as the energy scale around which localization effects appear. It is worth noticing that  
$s_c$ underestimates the spin correlation length, hence the actual localization length. 
The reason is that the RG equations blow up on a scale which is related to the largest gap in 
the excitation spectrum. Since the coupling $g$ 
only affects charge degrees of freedom, the largest gap is expected in  
the charge sector, the spin gap being smaller. Keeping this in mind,        
in what follows we shall discuss how $s_c$, or better $E_c$, depend on the range of the 
disorder potential. 
  
First of all we need to identify some reference scale to compare with $E_c$. 
The natural candidate would be the inverse relaxation time $1/\tau_0$ in the Born 
approximation. In the generic case $u^2\geq w^2\geq v^2$, $1/\tau_0=2Q_0$ where 
$Q_0$ is obtained by Eq.~(\ref{saddle-point}) with $u^2$ substituted by 
$u^2+2w^2+v^2$. We expect that the actual $E_c$ is always smaller then $1/\tau_0$, 
the two values being closest for extremely short-range disorder, namely 
$u^2=w^2=v^2$. Since the derivation of the 
non-linear $\sigma$-model does not provide with the precise dependence of 
the initial values of the coupling constants, $g^{(0)}$, $g_3^{(0)}$ 
and $g_+^{(0)}$, on the impurity potential, we will assume 
that the short range disorder 
corresponds to $2g^{(0)}=g_3^{(0)}=g_+^{(0)}$ and moreover 
that the mass gap $E_c$ in this case can be identified as $1/\tau_0$ in the Born approximation. 
Upon integrating the RG equations (\ref{dg}-\ref{dg+}) with this initial condition, 
the value of $E_c\simeq 1/\tau_0$ is found to be  
\be
\frac{1}{\tau_0} \simeq  \Lambda\, \exp\left(-\frac{1}{2g^{(0)}+g_3^{(0)}+g_+^{(0)}}\right),
\label{Ecsr}
\ee
where $\Lambda$ is the ultraviolet cut-off of the order of the gap $\Delta$. 
In order to appreciate the role of the range of the disorder-potential, 
let us analyze the RG equations (\ref{dg}-\ref{dg+}) keeping fixed the combination
$g_0=2g^{(0)}+g_3^{(0)}+g_+^{(0)}$, i.e. at constant $1/\tau_0$, and increasing the 
value of $g^{(0)}$ at expenses of $g_3^{(0)}+g_+^{(0)} = g_0-2g^{(0)}$. One readily realizes 
that as $g^{(0)}$ increases $E_c$ decreases from its short-range value (\ref{Ecsr}).  
For instance, if we assume 
$g_3^{(0)}=0$ and $g_+^{(0)} = g_0 - 2g^{(0)}\ll g_0$, then  
\[
E_c \simeq \frac{1}{\tau_0}\,\left(\frac{e g_+^{(0)}}{4g_0}\right)^{\frac{1}{g_0}}
\ll \frac{1}{\tau_0},
\]
which explicitly shows that localization effects may show up at energies/temperatures 
much smaller than the $1/\tau_0$ in the Born approximation.

\section{Conclusions}

In this work we have presented a derivation of the non-linear 
$\sigma$-model for disordered $d$-wave superconductors able to deal with a finite range 
impurity potential, namely with an intra-node scattering potential generically different 
from the inter-nodal one. Within this derivation we have been able to clarify 
some controversial issues concerning the validity of a conventional non-linear $\sigma$-model 
approach when dealing with disordered Dirac fermions. 
We have indeed found that the non-linear 
$\sigma$-model approach is actually equivalent to the alternative method first introduced in 
Ref.~\onlinecite{Nersesyan} which consists in mapping the disordered model onto a 
one-dimensional model of interacting fermions. 
The energy upper cut-off is provided by the inverse Born 
relaxation time $1/\tau_0$ in the non-linear $\sigma$-model approach and by the 
superconducting gap $\Delta$ in the 1d mapping. 
A closely related aspect 
which also emerges for $d$-wave superconductors 
is the existence of a Wess-Zumino-Novikov-Witten term related to the vorticity of the 
spectrum in momentum space.  
Both these features are responsible of several interesting phenomena. 
For instance, unlike conventional 
disordered systems, in this case disorder starts playing a role (particularly
in the density of states) when quasiparticle motion is still ballistic.  In contrast, 
the on-set of localization precursor effects is pushed towards energies/temperatures 
lower than the inverse relaxation time in the Drude approximation. More specifically, 
the longer is the range of the impurity potential the later localization effects appear. 
This result may explain why experiments have so often failed to detect localization precursor 
effects in cuprates superconductors. 
\\

{\bf Acknowledgments}: 
LD acknowledges support from the SFB TR 12 of the DFG and the EU network HYSWITCH.


\begin{thebibliography}{30}
\bibitem{coupling} For intra- and inter-opposite-node scattering, the non-linear $\sigma$-model 
coupling is instead of order one irrespectively of the velocity anisotropy, see the text below.
\end{thebibliography}
\end{document}